\documentclass[fleqn,usenatbib]{mnras}

\usepackage{newtxtext,newtxmath}
\usepackage[T1]{fontenc}

\DeclareRobustCommand{\VAN}[3]{#2}
\let\VANthebibliography\thebibliography
\def\thebibliography{\DeclareRobustCommand{\VAN}[3]{##3}\VANthebibliography}

\usepackage{graphicx}	% Including figure files
\usepackage{amsmath}	% Advanced maths commands

\title[Web--Halo Model in $f(R)$ gravity]
{Modelling the nonlinear matter power spectrum in Hu--Sawicki
$f(R)$ gravity with the Web--Halo Model}

\author[X. Wang \& Z. Huang]{
Xin Wang,$^{1}$\thanks{E-mail: wangx898@mail2.sysu.edu.cn}
and Zhiqi Huang$^{1,2}$\thanks{E-mail: huangzhq25@mail.sysu.edu.cn}
\\
$^{1}$School of Physics and Astronomy, Sun Yat-sen University,
Zhuhai 519082, P. R. China\\
$^{2}$CSST Science Center for the Guangdong-Hongkong-Macau Greater Bay Area,
Sun Yat-sen University, Zhuhai 519082, P. R. China
}

\date{Accepted XXX. Received YYY; in original form ZZZ}

\pubyear{\the\year{}}

\begin{document}
\label{firstpage}
\pagerange{\pageref{firstpage}--\pageref{lastpage}}
\maketitle

% Abstract of the paper
\begin{abstract}
We develop a semi-analytic extension of the Web--Halo Model (WHM) to Hu--Sawicki $f(R)$ gravity, with the aim of linking nonlinear matter clustering to the successive stages of cosmic-web collapse. The cylindrical sheet and filament windows of the original WHM are replaced by axisymmetric ellipsoidal top-hat windows, yielding a modest improvement around the perturbative-to-nonlinear transition without introducing additional fitting parameters. Modified gravity is incorporated through environment-dependent chameleon spherical collapse, from which the collapse threshold and virial overdensity are obtained and propagated through the sheet, filament, and halo contributions. For $|f_{R0}|=10^{-5}$ and $10^{-6}$, the predicted nonlinear enhancement is broadly consistent with the scale-, redshift-, and field-strength dependence seen in the e-MANTIS emulator, with closer agreement for the weaker field and at higher redshift. The component responses show that sheets and filaments contribute substantially around the transition regime, while the halo response becomes increasingly important at smaller scales. As complementary diagnostics, no displacement of the BAO peak is resolved, whereas tomographic weak-lensing spectra retain a percent-level response to the modified matter power spectrum. The extended WHM therefore provides a physically interpretable framework for tracing screened modified-gravity effects across the cosmic-web collapse hierarchy.
\end{abstract}

% Select between one and six entries from the list of approved keywords.
% Don't make up new ones.

%%%%%%%%%%%%%%%%%%%%%%%%%%%%%%%%%%%%%%%%%%%%%%%%%%

%%%%%%%%%%%%%%%%% BODY OF PAPER %%%%%%%%%%%%%%%%%%

\section{Introduction}
\label{sec:introduction}

The matter power spectrum is a fundamental summary statistic of large-scale structure. It enters the modelling of galaxy clustering, weak gravitational lensing, cluster abundance, and their cross-correlations, and therefore plays a central role in extracting cosmological information from current and forthcoming surveys. While linear theory provides an accurate description on sufficiently large scales, much of the statistical power of late-time structure probes comes from the quasi-nonlinear and nonlinear regimes. Reliable modelling of the nonlinear matter power spectrum is consequently a key requirement for modern cosmological analyses \citep{HUTERER_2005,Kilbinger_2015}.

This requirement has become increasingly important as weak-lensing and galaxy-clustering measurements have approached percent-level statistical precision. Analyses from DES \citep{Amon_2022,Secco_2022}, KiDS \citep{Asgari_2021}, and HSC \citep{Li_HSC_2023,Dalal_2023} are sensitive to both the amplitude and scale dependence of matter clustering, while requiring careful modelling of nonlinear structure, baryonic effects, intrinsic alignments, and redshift uncertainties. Spectroscopic surveys such as BOSS \citep{Alam_2017}, eBOSS \citep{Alam_2021}, and DESI \citep{Adame_2025} use BAO, redshift-space distortions, and full-shape clustering to constrain the expansion and growth histories. Future surveys, including Euclid \citep{Euclid_2020} and the Rubin Observatory LSST \citep{LSST_DESC_2021}, will place still greater demands on the accuracy and physical control of nonlinear predictions.

A variety of approaches have been developed to model the nonlinear matter power spectrum. Perturbative methods, including Eulerian and Lagrangian perturbation theory, provide controlled descriptions on large and mildly nonlinear scales and clarify the origin of mode coupling \citep{Bernardeau_2002,Crocce_2006,Matsubara_2008}. In the Lagrangian picture, the Zel'dovich approximation \citep{Zeldovich:1969sb} gives the first-order displacement solution and already captures the anisotropic nature of gravitational collapse before shell crossing, providing an intuitive connection between perturbation theory and the formation of sheet-like and filamentary cosmic-web structures \citep{Shandarin_1989,Hidding_2014}. More recently, the effective field theory of large-scale structure has extended the perturbative description by absorbing the effect of short-scale nonlinearities into counterterms, improving the reach of analytic predictions in the weakly nonlinear regime \citep{Carrasco_2012,Senatore_2015}.

On smaller scales, the most direct route is provided by cosmological $N$-body simulations and simulation-calibrated fitting formulae. High-resolution simulations can follow deeply nonlinear gravitational clustering, but they are computationally expensive, especially when they must be repeated over broad cosmological or modified-gravity parameter spaces. This has motivated fitting prescriptions such as HALOFIT \citep{Smith_2003,Takahashi_2012} and fast emulators such as CosmicEmu \citep{Heitmann_2016}, EuclidEmulator \citep{Knabenhans_2021}, and baccoemu \citep{Angulo_2021,Arico_2021}. Recently, machine-learning and neural-network emulators have also been developed to accelerate predictions for nonlinear statistics across extended parameter spaces \citep{Agarwal_2014,Yang_2026,Tanaka_2026}. These approaches can achieve high numerical accuracy, but the physical interpretation of the predicted nonlinear response is not always transparent because it is encoded in simulation-calibrated parameters or trained interpolation schemes.

The halo model provides a complementary semi-analytic description. It organizes nonlinear clustering in terms of virialized dark-matter haloes, combining ingredients such as the halo mass function, halo bias, density profiles, and concentration--mass relations \citep{Seljak_2000,COORAY_2002,Asgari_2023}. This framework connects the matter power spectrum to the statistics and internal structure of collapsed objects and has been widely used in both analytic modelling and phenomenological extensions. Modern implementations, including HMcode, improve the accuracy of the halo-model predictions through simulation calibration and phenomenological parameters, with extensions that account for baryonic feedback and cosmologies beyond minimal $\Lambda$CDM \citep{Mead_2015,Mead_2016_2,Mead_2021}.

Despite its usefulness, the standard halo model has well-known limitations. In particular, the transition between the perturbative regime and the halo-dominated regime is difficult to model accurately \citep{Acuto_2021,Mead_2021}. This is the range where neither purely perturbative calculations nor the one-halo term alone gives a complete physical description. In the actual cosmic web, matter does not collapse directly from the linear field into virialized haloes. Gravitational collapse is generically anisotropic: matter first forms sheets, then filaments, and finally haloes. This hierarchy is already present in the Zel'dovich approximation and subsequent descriptions of the cosmic web \citep{Zeldovich:1969sb,Shandarin_1989,Bond_1996,Shen_2006,Cautun_2014}. It is also supported by ellipsoidal-collapse models, in which the collapse threshold depends on the tidal shear and differs from the spherical-collapse prediction \citep{Bond_Myers_1996,Sheth_2001,Sheth_2002,Angrick_2010,Ludlow_2014}. A model based only on virialized haloes can therefore omit part of the physical structure that controls the transition from large-scale perturbative clustering to small-scale halo clustering.

The Web--Halo Model (WHM) was introduced to incorporate these
intermediate stages of anisotropic collapse \citep{Brieden_2026}. In addition to fully collapsed haloes, it includes structures that have collapsed along one dimension, identified with sheets, and along two dimensions, identified with filaments. The nonlinear matter power spectrum is written as a perturbative contribution supplemented by compensated sheet, filament, and halo corrections. These collapse orders do not represent independent matter components; instead, they describe a hierarchical redistribution of power as matter progresses through successive stages of anisotropic collapse. The WHM therefore provides a cosmic-web interpretation of the transition between perturbative and halo-dominated clustering.

A central feature of the WHM is that it improves the transition-regime prediction without introducing new fitting parameters. The original work showed that, by combining one-loop Lagrangian perturbation theory with compensated sheet, filament, and halo terms, the WHM can match state-of-the-art emulator predictions in the transition region with an accuracy competitive with calibrated halo-model prescriptions, while retaining a parameter-free structure \citep{Brieden_2026}. This makes the WHM an attractive framework not only for modelling the matter power spectrum in GR, but also for investigating how new gravitational dynamics propagate through different stages of cosmic-web collapse. More recently, the WHM hierarchy has been extended to halo bias, stochasticity, and assembly bias through a peak--background-split construction \citep{Brieden_Tipp_2026}.

Modified gravity provides a natural setting for a complementary extension of the WHM. Although the late-time accelerated expansion of the Universe is successfully described by a cosmological constant within $\Lambda$CDM, its physical origin remains unresolved. Modified-gravity theories offer an alternative possibility: cosmic acceleration, or deviations in the growth of structure, may arise from modifications of gravity on cosmological scales \citep{Clifton_2012,Joyce_2016,Koyama_2016}. Even when the background evolution remains close to $\Lambda$CDM, changes in the growth of structure can leave observable signatures in large-scale-structure statistics.

Among modified-gravity models, $f(R)$ gravity is one of the simplest and most widely studied examples \citep{Sotiriou_2010,De_Felice_2010}. The Hu--Sawicki model \citep{Hu_2007} introduces an additional scalar degree of freedom that mediates a fifth force. The chameleon mechanism suppresses this force in dense or deep-potential regions, allowing the model to satisfy local gravity constraints while retaining potentially observable cosmological effects \citep{Khoury_2004,Burrage_2018}. Nonlinear clustering in this model therefore depends not only on scale and redshift, but also on the mass, density, and environment of collapsing structures.

Nonlinear structure formation in modified gravity has been studied using perturbative calculations, spherical-collapse models, numerical simulations, and simulation-calibrated prescriptions. These studies have characterized screening signatures in the nonlinear power spectrum \citep{Koyama_2009,Cui_2010}, modified collapse thresholds and halo mass functions \citep{Lombriser_2013,Kopp_2013}, and changes in halo abundances, concentrations, and clustering \citep{Hagstotz_2019,Mitchell_2019,Mead_2016_2,Gupta_2023}. Dedicated modified-gravity emulators such as FORGE, e-MANTIS, and FREmu provide fast predictions for the nonlinear matter power spectrum or its boost relative to GR \citep{Arnold_2022,S_ez_Casares_2023,Bai_2024}. We use e-MANTIS as the numerical benchmark in this work because it emulates the nonlinear $f(R)$ boost, rather than only the absolute power spectrum, over the range of field strengths and redshifts relevant to our analysis.

These approaches have significantly advanced nonlinear modified-gravity modelling. However, many of them rely on simulation-calibrated fitting functions, emulator interpolation, or phenomenological modifications of halo-model ingredients. Such methods are powerful for precision prediction, but they do not always expose how modified collapse dynamics propagate through the different stages of structure formation. The WHM offers a complementary route: it provides a nonlinear power-spectrum prediction while allowing the response to be decomposed into sheet, filament, and halo corrections.

This motivates the present work. We aim to construct a semi-analytic extension of the WHM to Hu--Sawicki $f(R)$ gravity while preserving the physically motivated structure of the original model. Rather than applying a phenomenological boost to the final power spectrum, we introduce the modified-gravity effect through collapse dynamics. The chameleon-screened fifth force changes the spherical-collapse threshold and virial overdensity, and these quantities are then propagated into the sheet, filament, and halo contributions of the WHM.

We introduce two modifications to the WHM framework. First, the cylindrical sheet and filament windows are replaced by axisymmetric ellipsoidal windows. This new geometry preserves the original mass-conservation and overdensity assignments, introduces no additional fitting parameters, and is more closely aligned with anisotropic collapse. Second, we compute environment-dependent collapse quantities $\delta_{\rm c}^{f(R)}$ and $\Delta_{\rm v}^{f(R)}$ for the Hu--Sawicki model using a chameleon-screened thin-shell prescription. The resulting matter power spectrum is obtained by averaging the environment-dependent WHM prediction over the environmental density distribution.

We test the resulting framework in several ways. After assessing the effect of the ellipsoidal windows in GR, we compare the $f(R)$ WHM boost with the e-MANTIS emulator, decompose the modified-gravity response into sheet, filament, and halo contributions, and propagate the predicted matter power spectrum into BAO and weak-lensing diagnostics. The purpose is not to replace simulation-calibrated emulators, but to develop a physically interpretable framework that connects the nonlinear modified-gravity response to successive stages of cosmic-web collapse.

This paper is organized as follows. Section~\ref{sec:theory} reviews the WHM and the Hu--Sawicki $f(R)$ model. Section~\ref{sec:extensions} introduces the ellipsoidal window functions and the environment-dependent $f(R)$ collapse calculation, and describes their implementation within the WHM. Section~\ref{sec:results} presents the GR window-function test, the $f(R)$ matter-power-spectrum boost, its component decomposition, and the BAO and weak-lensing diagnostics. We discuss the results and summarize our conclusions in Section~\ref{sec:discussion_conclusions}.

\section{Theoretical background}
\label{sec:theory}

\subsection{Brief review of the Web--Halo Model}

For completeness, we summarize the WHM ingredients required below. Suppressing the explicit redshift dependence, the nonlinear matter power spectrum is constructed as
\begin{equation}
 P_{\rm WHM}(k)=P_{\rm PT}(k)+\hat{P}_{\rm 1s}(k)+\hat{P}_{\rm 1f}(k)+\hat{P}_{\rm 1h}(k),
\end{equation}
where $P_{\rm PT}$ denotes the perturbative prediction and determines the large-scale limit, while $\hat{P}_{\rm 1s}, \hat{P}_{\rm 1f}$ and $\hat{P}_{\rm 1h}$ are the compensated sheet-, filament-, and halo-order corrections, respectively. Formally, they are written as
\begin{align}
    \hat{P}_{\rm 1s}(k)&=\frac{1}{\bar{\rho}}\int_0^\infty M(\nu)\left[W_{\rm s}^2(k,\nu)-W_{\rm PT}^2(k,\nu)\right]f_{\rm s}(\nu)\,\mathrm{d}\nu, \\
    \hat{P}_{\rm 1f}(k)&=\frac{1}{\bar{\rho}}\int_0^\infty M(\nu)\left[W_{\rm f}^2(k,\nu)-W_{\rm s}^2(k,\nu)\right]f_{\rm f}(\nu)\,\mathrm{d}\nu, \\
    \hat{P}_{\rm 1h}(k)&=\frac{1}{\bar{\rho}}\int_0^\infty M(\nu)\left[W_{\rm h}^2(k,\nu)-W_{\rm f}^2(k,\nu)\right]f_{\rm h}(\nu)\,\mathrm{d}\nu .
\end{align}
Here $\bar\rho$ is the mean comoving matter density, $W_{\rm s}, W_{\rm f}$, and $W_{\rm h}$ are the window functions associated with sheets, filaments, and haloes, while $f_{\rm s}, f_{\rm f}$, and $f_{\rm h}$ are the corresponding multiplicity functions. The function $W_{\rm PT}$ denotes the perturbative window before the first shell crossing. The integration variable $\nu=\delta_{\rm c}/\sigma(R)$ is the peak height, where $\delta_{\rm c}$ is the linear overdensity threshold for spherical collapse and $\sigma(R)$ is the variance of the linear density field smoothed on the Lagrangian radius $R(M)=(3M/4\pi\bar{\rho})^{1/3}$.

The hats emphasize that the one-object terms are compensated corrections rather than independent additive components of the matter field. The WHM adopts a hierarchical substructure picture: haloes are embedded in parent filaments, and filaments are embedded in parent sheets. Progression from one collapse order to the next therefore redistributes the power associated with the same mass instead of adding a new population of matter. This is implemented by subtracting the window function of the previous order: the sheet correction subtracts the perturbative window, the filament correction subtracts the sheet window, and the halo correction subtracts the filament window. Since the normalized windows satisfy $W_X(k\rightarrow0)\rightarrow1$, each
compensated correction vanishes in the large-scale limit. This prevents double counting and implements mass conservation across the collapse hierarchy.

\subsection{Hu--Sawicki \texorpdfstring{$f(R)$}{f(R)} gravity}

In $f(R)$ gravity, the Einstein--Hilbert action is generalized to 
\begin{equation}
S=\int \mathrm{d}^4x\sqrt{-g}\left[\frac{R+f(R)}{16\pi G}+\mathcal{L}_{\rm m}\right],
\end{equation}
where $R$ is the Ricci scalar and $\mathcal{L}_{\rm m}$ is the matter Lagrangian. The derivative $f_R\equiv \mathrm{d}f/\mathrm{d}R$ acts as an additional scalar degree of freedom. This scalar field mediates a fifth force and can therefore modify gravitational clustering unless it is suppressed by a screening mechanism.

In this work, we focus on the Hu--Sawicki model \citep{Hu_2007}, for which the function $f(R)$ can be written as
\begin{equation}
f(R)=-m^2\frac{c_1(R/m^2)^n}{c_2(R/m^2)^n+1},
\end{equation}
where $m^2=H_0^2\Omega_m$. The model parameters can be chosen such that the leading constant term reproduces an effective cosmological constant and the background expansion closely follows that of $\Lambda$CDM. In the high-curvature regime, $R\gg m^2$, the model is commonly parameterized by the present background scalar field value $f_{R0}\equiv f_R(\bar{R}_0)$, giving
\begin{equation}
f(R) \simeq -2\Lambda - \frac{f_{R0}}{n}\frac{\bar{R}_0^{n+1}}{R^n},
\end{equation}
where $\bar{R}_0$ denotes the present background curvature. The corresponding scalar field is
\begin{equation}
f_R(R)\simeq f_{R0}\left(\frac{\bar R_0}{R}\right)^{n+1}.
\end{equation}
Throughout this work, we follow the convention in the literature and fix the index $n=1$, so that the strength of the modification is characterized by $|f_{R0}|$.

A key feature of Hu--Sawicki $f(R)$ gravity is the chameleon screening mechanism. In the high-curvature regime, the effective scalar-field mass is approximately
\begin{equation}
m_{\rm eff}^2\simeq\frac{1}{3f_{RR}},
\end{equation}
where $f_{RR}=\mathrm{d}^2f/\mathrm{d}R^2$. In dense or deep-potential regions, the large curvature increases the effective mass and shortens the range of the scalar-mediated interaction, suppressing the modification to gravity. In lower-density or shallower-potential regions, the field can remain sufficiently light to mediate an additional attractive force. The resulting force enhancement therefore depends not only on time and scale, but also on the density of the collapsing object and its environment.

The collapse dynamics provide a natural interface between this screened force and the WHM. In particular, the collapse threshold $\delta_{\rm c}$ enters the peak height $\nu$, while the virial overdensity $\Delta_{\rm v}$ determines the structural scales assigned to the different collapse orders. We therefore incorporate the modified-gravity response through an environment-dependent collapse calculation, from which $\delta_{\rm c}$ and $\Delta_{\rm v}$ are derived. The implementation of this construction is described in Section~\ref{sec:extensions}.

\section{Extensions of the WHM framework}
\label{sec:extensions}

\subsection{Ellipsoidal window functions}
\label{sec:ellipsoidal_window}

In the original WHM, the sheet and filament window functions are modelled using idealized cylindrical geometries. This choice leads to simple analytic expressions and introduces no fitting parameters. However, sheets and filaments arise from anisotropic collapse, and their geometry is not expected to be well represented by cylinders. Since the window functions determine how the mass associated with each collapse order contributes to the power spectrum, their geometrical form can affect the transition between the perturbative and halo-dominated regimes.

We therefore replace the cylindrical approximation by axisymmetric ellipsoidal windows. This is a geometrical refinement of the WHM ingredients rather than a phenomenological correction to the final power spectrum. The ellipsoidal semi-axes are fixed by the same mass-conservation and overdensity assignments used in the original WHM, so no additional fitting parameters are introduced.

Consider a homogeneous axisymmetric ellipsoid with transverse semi-axis $a_\perp$ and longitudinal semi-axis $a_\parallel$. Its normalized density profile is
\begin{equation}
\frac{\rho_{\rm ell}(\mathbf r)}{M}=\frac{3}{4\pi a_\perp^2a_\parallel}\Theta\left(1-\frac{x^2+y^2}{a_\perp^2}-\frac{z^2}{a_\parallel^2}\right),
\end{equation}
where $\Theta$ is the Heaviside step function. The normalized Fourier-space window is
\begin{align}
W_{\rm ell}(\mathbf k)&=\int\frac{\rho_{\rm ell}(\mathbf r)}{M}e^{i\mathbf k\cdot\mathbf r}\,d^3\mathbf r\nonumber\\
&=W_{\rm t}(Q)=3\frac{\sin Q-Q\cos Q}{Q^3},
\end{align}
where
\begin{equation}
  Q^2=k^2\left[a_\perp^2(1-\mu^2)+a_\parallel^2\mu^2\right].
\end{equation}
Here $\mu$ is the cosine of the angle between the wave vector $\mathbf{k}$ and the symmetry axis of the ellipsoid. This expression follows from a linear coordinate transformation that maps the ellipsoid to a sphere. In the spherical limit $a_\perp=a_\parallel=R$, it reduces to the usual top-hat window $W_{\rm t}(kR)$.

Since the matter power spectrum is statistically isotropic, the quantity entering the WHM integrals is the orientation average of the squared window. We define
\begin{equation}
  W_{\rm ell}^2(k)=\int_0^1 \mathrm{d}\mu\,W_{\rm t}^2\left[k\sqrt{a_\perp^2(1-\mu^2)+a_\parallel^2\mu^2}\right].
\end{equation}
The integration range $0\leq\mu\leq1$ is equivalent to the usual angular average $(1/2)\int_{-1}^{1}\mathrm{d}\mu$, because the squared window is symmetric under $\mu\rightarrow-\mu$.

The ellipsoidal semi-axes are assigned in direct analogy with the cylindrical construction of the original WHM. Mass conservation requires
\begin{equation}
  M=\bar\rho\,\frac{4\pi}{3}R^3=\Delta_X\bar\rho\,V_X ,
\end{equation}
where $X=s,f$ labels the sheet and filament collapse orders. In the original cylindrical prescription, $V_X=\pi a^2 b$. For an
axisymmetric ellipsoid, the corresponding volume is
\begin{equation}
  V_X^{\rm ell}=\frac{4\pi}{3}a_{\perp,X}^2a_{\parallel,X}.
\end{equation}
We retain the original WHM overdensity assignment,
\begin{equation}
  \Delta_s=\Delta_{\rm v}^{1/3}, \qquad \Delta_f=\Delta_{\rm v}^{2/3}.
\end{equation}

For sheets, corresponding to collapse along one spatial dimension, the two uncollapsed transverse directions remain at the Lagrangian scale, while the symmetry-axis direction is compressed:
\begin{equation}
  a_{\perp,s}=R, \qquad a_{\parallel,s}=\frac{R}{\Delta_s}=\frac{R}{\Delta_{\rm v}^{1/3}}.
\end{equation}
This gives
\begin{equation}
  V_s^{\rm ell}=\frac{4\pi}{3}a_{\perp,s}^2a_{\parallel,s}=\frac{4\pi}{3}\frac{R^3}{\Delta_s}.
\end{equation}

For filaments, corresponding to collapse along two spatial dimensions, the two transverse directions are compressed, while the longitudinal symmetry-axis direction remains at the Lagrangian scale:
\begin{equation}
  a_{\perp,f}=\frac{R}{\sqrt{\Delta_f}}=\frac{R}{\Delta_{\rm v}^{1/3}}, \qquad a_{\parallel,f}=R.
\end{equation}
The corresponding volume is
\begin{equation}
  V_f^{\rm ell}=\frac{4\pi}{3}a_{\perp,f}^2a_{\parallel,f}=\frac{4\pi}{3}\frac{R^3}{\Delta_f}.
\end{equation}

Thus the ellipsoidal prescription preserves the same mass and effective overdensity assignments as the original WHM cylindrical model, while modifying only the geometrical shape of the sheet and filament windows. The halo contribution is left unchanged and is still described by the spherical NFW window.

The motivation for this replacement is twofold. From a theoretical point of view, the WHM is based on anisotropic collapse, and an ellipsoidal window provides a more direct geometrical representation of a Lagrangian region compressed along one or two principal axes. From a practical point of view, the ellipsoidal window has a compact form: the usual top-hat argument $kR$ is simply replaced by the ellipsoidal variable $Q$. This avoids the approximate cylindrical-window expressions used in the original WHM, while preserving the original mass-conservation conditions.

The replacement therefore changes only the sheet and filament geometries. Its impact is expected to be concentrated in the transition between the perturbative and halo-dominated regimes, while the perturbative contribution and spherical NFW halo window remain unchanged.

\subsection{\texorpdfstring{$f(R)$}{f(R)} modification of the collapse dynamics}

The modified-gravity dependence is introduced through the dynamics of spherical collapse rather than through a phenomenological correction to the final WHM power spectrum. The chameleon-screened fifth force changes the nonlinear evolution of a collapsing top-hat perturbation, from which the collapse quantities $\delta_{\rm c}$ and $\Delta_{\rm v}$ entering the WHM are derived.

\subsubsection{Effective force enhancement}
\label{effective_force_enhancement}

We describe the modified gravitational force in terms of an effective gravitational constant,
\begin{equation}
G_{\rm eff}=G(1+F),
\end{equation}
where $F$ is the chameleon-screened enhancement of gravity. It interpolates between the fully screened limit $F=0$ and the fully unscreened limit $F=1/3$.

Following \citet{Khoury_2004,Lombriser_2013}, we use the thin-shell approximation to model the fifth-force enhancement around a spherical top-hat region. The physical top-hat radius is
\begin{equation}
R_{\rm TH}(a)=\frac{aR}{(1+\delta)^{1/3}},
\end{equation}
where $R$ is the comoving Lagrangian radius and $\delta$ is the nonlinear density contrast inside the top-hat. This relation follows from conservation of the mass enclosed by the collapsing region.

For the $n=1$ Hu--Sawicki model adopted in this work, the thin-shell factor is written as
\begin{align}
\frac{\Delta R}{R_{\rm TH}}&=\frac{|f_{R0}|a^3}{\Omega_m\widetilde{\rho}_{\rm in}(H_0R_{\rm TH}/c)^2}\Bigg[\left(
\frac{1+4\Omega_\Lambda/\Omega_m}{\widetilde{\rho}_{\rm out}a^{-3}+4\Omega_\Lambda/\Omega_m}\right)^2
\nonumber\\
&\hspace{2.5em}-\left(\frac{1+4\Omega_\Lambda/\Omega_m}{\widetilde{\rho}_{\rm in}a^{-3}+4\Omega_\Lambda/\Omega_m}\right)^2\Bigg],
\label{eq:thin_shell_factor}\\
F&=\frac13\min\left[3\frac{\Delta R}{R_{\rm TH}}-3\left(\frac{\Delta R}{R_{\rm TH}}\right)^2+\left(\frac{\Delta R}{R_{\rm TH}}\right)^3,1\right],
\label{eq:F_full}
\end{align}
where $a$ is the scale factor, $c$ is the speed of light, $H_0$ is the Hubble constant, and $\Omega_m$ and $\Omega_{\Lambda}$ are the present-day matter and cosmological-constant density parameters. The dimensionless densities entering the thin-shell estimator are
\begin{equation}
\widetilde{\rho}_{\rm in}=1+\delta,\qquad\widetilde{\rho}_{\rm out}=1+\delta_{\rm env}^{\rm NL}.
\end{equation}
These quantities represent the matter density inside the sphere and in its environment, respectively, in units of the background matter density. The environmental density contrast $\delta_{\rm env}^{\rm NL}$ will be discussed in Section~\ref{sec:environmental_averaging}. 

In the geometrical thin-shell picture,
$\Delta R=R_{\rm TH}-R_0$, where $R_0$ is the radius of the screened interior. A physical thin shell therefore satisfies $0\leq\Delta R/R_{\rm TH}\leq1$, corresponding to $0\leq F\leq1/3$. The two limits describe a completely screened object and a fully unscreened object, respectively.

For the numerical calculations, we adopt the simplified interpolation
\begin{equation}
F=\frac{1}{3}\min\left[\max\left(3\alpha\frac{\Delta R}{R_{\rm TH}},0\right),1\right],
\label{eq:F_used}
\end{equation}
with the fixed coefficient $\alpha=1/2$. This prescription has been used in previous studies to account effectively for the difference between the idealized top-hat thin-shell approximation and the behaviour measured in $f(R)$ simulations \citep{Li_2012,Lombriser_2013}. We therefore adopt it as our fiducial prescription for the chameleon-screened enhancement factor. The upper bound gives the fully unscreened limit $F=1/3$, while the lower clipping enforces the physical bound $F\geq0$. In particular, the formal thin-shell estimator can become negative when $\widetilde{\rho}_{\rm out}>\widetilde{\rho}_{\rm in}$, in which case the lower bound sets $F=0$.

\subsubsection{Nonlinear and linear collapse equations}

We introduce the $f(R)$ effect by modifying the nonlinear evolution equation of the density contrast. In the spherical-collapse equation, the source term on the right-hand side originates from the Poisson equation and is therefore proportional to the gravitational constant. We incorporate the fifth force by replacing $G$ with the effective coupling $G_{\rm eff}=G(1+F)$. The density contrast therefore evolves according to
\begin{equation}
\delta''+\left(\frac{3}{a}+\frac{H'}{H}\right)\delta'-\frac{4}{3}\frac{\delta'^2}{1+\delta}=\frac{3}{2}\frac{\Omega_m(a)}{a^2}\delta(1+\delta)(1+F),
\label{eq:nonlinear_collapse}
\end{equation}
where a prime denotes differentiation with respect to the scale factor, $\mathrm{d}/\mathrm{d}a$, and $H$ is the Hubble parameter. The enhancement factor is evaluated using equation~\eqref{eq:F_used} and depends on the Lagrangian scale, time, nonlinear density contrast, and environment,
\begin{equation}
    F=F(R,a,\delta,\delta_{\rm env}^{\rm NL}).
\end{equation}
As a result, the collapse solution depends on the mass scale, collapse redshift, and environmental density.

To define the linearly extrapolated collapse threshold, we also construct a linear evolution equation associated with the same real-space screening prescription. We retain only terms that are first order in the top-hat density contrast and evaluate the enhancement factor at $\delta=0$. The resulting equation is
\begin{equation}
\delta_{\rm lin}''+\left(\frac{3}{a}+\frac{H'}{H}\right)\delta_{\rm lin}'=\frac{3}{2}\frac{\Omega_m(a)}{a^2}\delta_{\rm lin}(1+F_{\rm lin}),
\label{eq:linear_collapse}
\end{equation}
where
\begin{equation}
F_{\rm lin}=F(R,a,\delta=0,\delta_{\rm env}^{\rm NL}).
\end{equation}
Since $F_{\rm lin}$ is independent of $\delta_{\rm lin}$, this construction preserves the linearity of the evolution equation while maintaining a direct connection to the real-space screening model used for nonlinear collapse.

This associated linear equation should not be identified with the full Fourier-space linear response of $f(R)$ gravity. It is a practical real-space prescription introduced for the spherical-collapse construction. Setting $\delta=0$ fixes $\widetilde{\rho}_{\rm in}=1$, so that an overdense environment has $\widetilde{\rho}_{\rm out}>\widetilde{\rho}_{\rm in}$ and can drive the formal thin-shell estimator below its physical lower bound. The lower clipping in equation~\eqref{eq:F_used} then sets $F_{\rm lin}=0$. This produces a sharp but continuous environmental transition in $\delta_{\rm c}^{f(R)}$, which should be regarded as a limitation of the real-space linearization rather than as a physical non-smoothness of linear perturbations in $f(R)$ gravity.

\subsubsection{Spherical collapse quantities}

In the original WHM implementation, the spherical-collapse threshold $\delta_{\rm c}$ and the virial overdensity $\Delta_{\rm v}$ are taken from GR fitting functions \citep{Mead_2016}. In the present $f(R)$ extension, these inputs are instead computed from the modified collapse equations~\eqref{eq:nonlinear_collapse} and \eqref{eq:linear_collapse}. Thus $\delta_{\rm c}$ and $\Delta_{\rm v}$ are outputs of the environment-dependent collapse calculation rather than phenomenological modifications introduced at the level of the power spectrum.

We first determine the collapse threshold $\delta_{\rm c}^{f(R)}$. For a given Lagrangian radius $R$, target redshift $z$, and environment $\delta_{\rm env}$, we solve the nonlinear equation~\eqref{eq:nonlinear_collapse} using a shooting procedure analogous to that of \citet{Pace_2010}. The integration is initialized at $a_i=10^{-4}$, and the early-time growing-mode condition is imposed through
\begin{equation}
\delta_i'=\frac{\delta_i}{a_i},
\end{equation}
This relation corresponds to the early-time behaviour $\delta\propto a$ of the background model used in the collapse calculation.

We then vary the growing-mode amplitude $\delta_i$ until the nonlinear solution reaches $\delta(a_{\rm coll})=10^7$ at the target collapse scale factor $a_{\rm coll}=1/(1+z)$. The finite threshold $\delta=10^7$ is used as a numerical proxy for the formal divergence $\delta\rightarrow\infty$. Once the value of $\delta_i$ producing collapse at $a_{\rm coll}$ has been found, we solve the linear equation~\eqref{eq:linear_collapse} with the same initial pair,
\begin{equation}
\left(\delta_i,\delta_i'\right)=\left(\delta_i,\frac{\delta_i}{a_i}\right).
\end{equation}
The collapse threshold is then defined as
\begin{equation}
\delta_{\rm c}^{f(R)}(R,z,\delta_{\rm env})=\delta_{\rm lin}(R,a_{\rm coll},\delta_{\rm env}).
\end{equation}
Thus, the nonlinear shooting procedure determines the growing-mode initial amplitude required for collapse, while its linearly evolved value defines $\delta_{\rm c}^{f(R)}$.

We next compute the virial overdensity $\Delta_{\rm v}^{f(R)}$. In the spherical-collapse model, the turnaround epoch $a_{\rm ta}$ is identified as the time when the physical radius of the collapsing top-hat reaches its maximum. Since mass conservation gives
\begin{equation}
R_{\rm TH}(a) \propto \frac{a}{(1+\delta)^{1/3}},
\end{equation}
we determine $a_{\rm ta}$ by locating the maximum of $R_{\rm TH}(a)$ along the nonlinear collapse trajectory.

After turnaround, the perturbation is assumed to virialize at
\begin{equation}
r_{\rm vir}=s_{\rm vir} r_{\rm ta},
\end{equation}
where the virial ratio $s_{\rm vir}$ is determined from the
$\Lambda$CDM virial-closure equation
\begin{equation}
2\eta_\Lambda s_{\rm vir}^3-(2+\eta_\Lambda)s_{\rm vir}+1=0,
\label{eq:svir_equation_main}
\end{equation}
with
\begin{equation}
\eta_\Lambda=2\frac{\Omega_\Lambda}{\Omega_m}\frac{a_{\rm ta}^3}{1+\delta(a_{\rm ta})}.
\end{equation}
We select the physical root that is continuously connected to $s_{\rm vir}=1/2$ in the Einstein--de Sitter limit $\eta_\Lambda\rightarrow0$.

The original WHM fitting prescription adopts the Einstein--de Sitter value $s_{\rm vir}=1/2$. In a $\Lambda$CDM background, however, the cosmological constant modifies the virialization condition and $s_{\rm vir}$ is no longer exactly one half. We therefore solve equation~\eqref{eq:svir_equation_main} separately for each collapse trajectory.

In $f(R)$ gravity the fifth force can in principle also modify the energy balance between turnaround and virialization. Appendix~\ref{app:virialization} shows that representing this effect by a constant effective enhancement $\bar F$ changes the physical root by only a few $\times10^{-3}$ for the representative cases considered there. We therefore adopt the $\bar F=0$ closure equation~\eqref{eq:svir_equation_main} in the fiducial calculation. The turnaround quantities entering this equation nevertheless remain those obtained from the modified nonlinear collapse trajectory.

With this prescription, the matter density at virialization is
\begin{equation}
\rho_{m,\rm vir}=\frac{M}{\frac{4}{3}\pi r_{\rm vir}^3}=\frac{1}{s_{\rm vir}^3}\frac{M}{\frac{4}{3}\pi r_{\rm ta}^3}=\frac{1}{s_{\rm vir}^3}\rho_m(a_{\rm ta}).
\end{equation}
Identifying the virialized object with the target collapse epoch $a_{\rm coll}$, the virial overdensity relative to the background matter density is therefore
\begin{align}
\Delta_{\rm v}&=\frac{\rho_{m,\rm vir}}{\bar{\rho}_m(a_{\rm coll})}=\frac{1}{s_{\rm vir}^3}\frac{\rho_m(a_{\rm ta})}{\bar{\rho}_m(a_{\rm coll})} =\frac{1}{s_{\rm vir}^3}\frac{\rho_m(a_{\rm ta})}{\bar{\rho}_m(a_{\rm ta})}\left(\frac{a_{\rm coll}}{a_{\rm ta}}\right)^3 \notag \\
&=\frac{1}{s_{\rm vir}^3}\left[1+\delta(a_{\rm ta})\right]\left(\frac{a_{\rm coll}}{a_{\rm ta}}\right)^3.
\label{eq:Delta_v_fR}
\end{align}
Because both $a_{\rm ta}$ and $\delta(a_{\rm ta})$ are obtained from the modified nonlinear evolution, the virial overdensity is also a function of scale, redshift, and environment,
\begin{equation}
\Delta_{\rm v}^{f(R)}(R,z,\delta_{\rm env})=\Delta_{\rm v}(R,a_{\rm coll},\delta_{\rm env}).
\end{equation}
Representative collapse outputs and a detailed diagnostic of the
clipping-induced environmental transition are presented in
Appendix~\ref{app:collapse_diagnostics}.

\subsubsection{Environment}
\label{sec:environmental_averaging}

The chameleon mechanism makes the spherical-collapse dynamics dependent on the density of the environment. In the thin-shell factor, equation~\eqref{eq:thin_shell_factor}, this dependence enters through $\widetilde{\rho}_{\rm out}=1+\delta_{\rm env}^{\rm NL}$. Here $\delta_{\rm env}^{\rm NL}$ denotes the nonlinear density contrast of the environment, while $\delta_{\rm env}$ denotes the corresponding linearly extrapolated density contrast. The latter is the variable used to describe the probability distribution of environments.

Following \citet{Lombriser_2013}, we assume that the environment itself evolves according to the $\Lambda$CDM spherical-collapse dynamics. In practice, the environmental trajectory is obtained from the GR form of equation~\eqref{eq:nonlinear_collapse}, i.e. with $F=0$. The resulting nonlinear environmental density contrast $\delta_{\rm env}^{\rm NL}$ is then used as the exterior density in the force-enhancement factor of the collapsing object.

For the numerical implementation, we first sample the linearly extrapolated environmental density on the finite interval
\begin{equation}
\delta_{\rm env}(z)\in \left[-n_{\rm env}\delta_{\rm c}^{\Lambda}(z),\delta_{\rm c}^{\Lambda}(z)\right),
\label{eq:environment_grid}
\end{equation}
where $\delta_{\rm c}^{\Lambda}$ is the spherical-collapse threshold in $\Lambda$CDM. We take $n_{\rm env}=3$. The upper limit ensures that the environment itself has not collapsed by the target redshift, while the lower limit truncates the far underdense tail for numerical calculation purposes.

For each value of $\delta_{\rm env}$ on this grid, we obtain the corresponding initial amplitude by extrapolating it back to the same initial scale factor $a_i$ using the GR linear growth factor,
\begin{equation}
\delta_{\rm env,i}=\delta_{\rm env}\frac{D(a_i)}{D(a_{\rm coll})}.
\end{equation}
The corresponding initial derivative is fixed by the same early-time growing-mode condition,
\begin{equation}
\delta_{\rm env,i}'=\delta_{\rm env,i}\frac{D'(a_i)}{D(a_i)}\simeq\frac{\delta_{\rm env,i}}{a_i},
\end{equation}
where the final approximation follows from $D(a)\propto a$ in the early-time limit of the background model used here. These initial conditions are then used to evolve the GR nonlinear spherical-collapse equation up to $a_{\rm coll}$. This produces the environmental trajectory $\delta_{\rm env}^{\rm NL}(a)$, which is interpolated and used in equation~\eqref{eq:thin_shell_factor} during the collapse calculation.

We adopt the Eulerian definition of the environment. In this prescription, the environment is specified by a fixed Eulerian radius $\zeta$, rather than by a fixed Lagrangian radius. This choice is physically motivated because the chameleon force is sensitive to the matter distribution over a finite spatial range. Following \citet{Li_2012_env}, we take $\zeta=5\,h^{-1}{\rm Mpc}$.

The approximate probability distribution of the linearly extrapolated Eulerian environment is
\begin{align}
P_\zeta(\delta_{\rm env})=&\frac{\beta^{\omega/2}}{\sqrt{2\pi}}\left[1+(\omega-1)\frac{\delta_{\rm env}}{\delta_{\rm c}^{\Lambda}}\right]\left(1-\frac{\delta_{\rm env}}{\delta_{\rm c}^{\Lambda}}\right)^{-\omega/2-1} \notag \\
&\times\exp\left[-\frac{\beta^\omega}{2}\frac{\delta_{\rm env}^2}{\left(1-\delta_{\rm env}/\delta_{\rm c}^{\Lambda}\right)^\omega} \right],
\label{eq:eulerian_environment_distribution}
\end{align}
where
\begin{equation}
\beta=\frac{\left[\zeta/(8\,h^{-1}{\rm Mpc})\right]^{3/\delta_{\rm c}^{\Lambda}}}{\sigma_8^{2/\omega}},\qquad\omega=\delta_{\rm c}^{\Lambda}\gamma.
\end{equation}
Here $\gamma=-\mathrm{d}\ln S/\mathrm{d}\ln M$ is the logarithmic slope of the density variance with respect to mass, with $S(M)=\sigma^2(M)$. For a power-law matter power spectrum $P(k)\propto k^{n_{\rm eff}}$, this reduces to $\gamma=(n_{\rm eff}+3)/3$. The distribution in equation~\eqref{eq:eulerian_environment_distribution} is used below to average the environment-dependent WHM prediction.

\subsection{\texorpdfstring{$f(R)$}{f(R)} WHM}
\label{sec:fR_WHM}

The modified collapse calculation yields two scale-, redshift-, and environment-dependent quantities,
\begin{equation}
\delta_{\rm c}^{f(R)}(R,z,\delta_{\rm env}),
\qquad
\Delta_{\rm v}^{f(R)}(R,z,\delta_{\rm env}).
\end{equation}
These quantities replace the corresponding GR collapse inputs in the WHM calculation. In particular, the peak-height variable becomes
\begin{equation}
\nu(R,z,\delta_{\rm env})=\frac{\delta_{\rm c}^{f(R)}(R,z,\delta_{\rm env})}{\sigma_{\rm GR}(R,z)},
\end{equation}
while the virial radius entering the halo and web-window functions is
\begin{equation}
r_{\rm v}(R,z,\delta_{\rm env})=\frac{R}{\left[\Delta_{\rm v}^{f(R)}(R,z,\delta_{\rm env})\right]^{1/3}}.
\end{equation}
In this work, the variance $\sigma(R,z)$ and the perturbative contribution $P_{\rm PT}$ are kept fixed to their GR predictions. We also retain the original functional forms of the WHM multiplicity functions and the concentration--mass relation. The multiplicity weights nevertheless acquire a modified-gravity dependence through the environment-dependent peak height $\nu$. These choices are made in order to isolate the response generated by the modified nonlinear collapse quantities.

For a fixed environment $\delta_{\rm env}$, the above replacements give the conditional WHM prediction,
\begin{equation}
P_{\rm WHM}^{f(R)}(k,z\mid\delta_{\rm env}).
\end{equation}
The final prediction of interest is not the distribution of this power spectrum over environments, but its environment-averaged value. This is analogous in spirit to the environmental averaging in excursion-set calculations, but here the average is applied directly to the WHM matter power spectrum rather than to the first-crossing distribution \citep{Lombriser_2013}.

Using the Eulerian environment probability distribution in equation~\eqref{eq:eulerian_environment_distribution}, we define
\begin{equation}
\left\langle P_{\rm WHM}^{f(R)} \right\rangle(k,z)=\int_{\delta_{\rm min}(z)}^{\delta_{\rm c}^{\Lambda}(z)}\mathrm{d}\delta_{\rm env}P_\zeta(\delta_{\rm env})P_{\rm WHM}^{f(R)}(k,z\mid\delta_{\rm env}),
\label{eq:environment_averaged_fR_WHM}
\end{equation}
where $\delta_{\rm min}(z)=-n_{\rm env}\delta_{\rm c}^{\Lambda}(z)$ is the lower bound of the finite environment grid in equation~\eqref{eq:environment_grid}. In the numerical calculation, $P_\zeta(\delta_{\rm env})$ is normalized over this interval before the average is evaluated.

The nonlinear power spectrum boost predicted by the modified WHM is then defined as
\begin{equation}
B_{\rm WHM}(k,z)=\frac{\left\langle P_{\rm WHM}^{f(R)} \right\rangle(k,z)}{P_{\rm WHM}^{\rm GR}(k,z)}.
\label{eq:WHM_boost}
\end{equation}
The GR denominator is evaluated using the same ellipsoidal-window WHM prescription and fiducial cosmology as the modified-gravity calculation. Since $F=0$ in the GR limit, the collapse quantities have no environmental dependence and no corresponding environmental average is required in the denominator.

Unless otherwise stated, all numerical results are obtained for a flat $\Lambda$CDM cosmology based on the Narya parameter set adopted in the original WHM calculation. We use
\begin{equation}
\begin{aligned}
\Omega_m &= 0.3613,\qquad
\Omega_b = 0.050,\qquad
h = 0.70,\\
n_s &= 1.01,\qquad
\sigma_8 = 0.90.
\end{aligned}
\end{equation}
We further set $w_0=-1$ and $w_a=0$, with $\Omega_\Lambda=1-\Omega_m$. Following the convention used in the original WHM implementation, the density corresponding to a fiducial neutrino mass sum $\sum m_\nu=0.06\,{\rm eV}$ is absorbed into the cold-matter density, while no massive-neutrino component is evolved explicitly. This accounts for the small difference between the value $\Omega_m=0.3613$ used here and the nominal Narya value $\Omega_m=0.360$. The same fiducial parameters are used throughout the WHM calculation and in the BACCO and e-MANTIS comparisons \citep{Angulo_2021,S_ez_Casares_2023,Brieden_2026}.

\section{Results}
\label{sec:results}

\subsection{Effect of the ellipsoidal window functions in GR}
\label{sec:ellipsoidal_window_results}

Before turning to the $f(R)$ results, we first examine the effect of replacing the original cylindrical sheet and filament windows by the ellipsoidal windows introduced in Section~\ref{sec:ellipsoidal_window}. This comparison is performed in GR, so that the impact of the geometric modification can be isolated from the modified-gravity effects. We compare both WHM predictions with the nonlinear matter power spectrum obtained from the BACCO emulator \citep{Angulo_2021}, which is used here as the GR nonlinear reference spectrum. We also mark the characteristic nonlinear scale $r_{\rm nl}$, defined as in the original WHM by
\begin{equation}
\sigma(r_{\rm nl}(z))=\delta_{\rm c}(z),
\label{eq:rnl_definition}
\end{equation}
and denote the corresponding wavenumber by $k_{\rm nl}(z)=r_{\rm nl}^{-1}(z)$.

\begin{figure*}
\centering
\includegraphics[width=\textwidth]{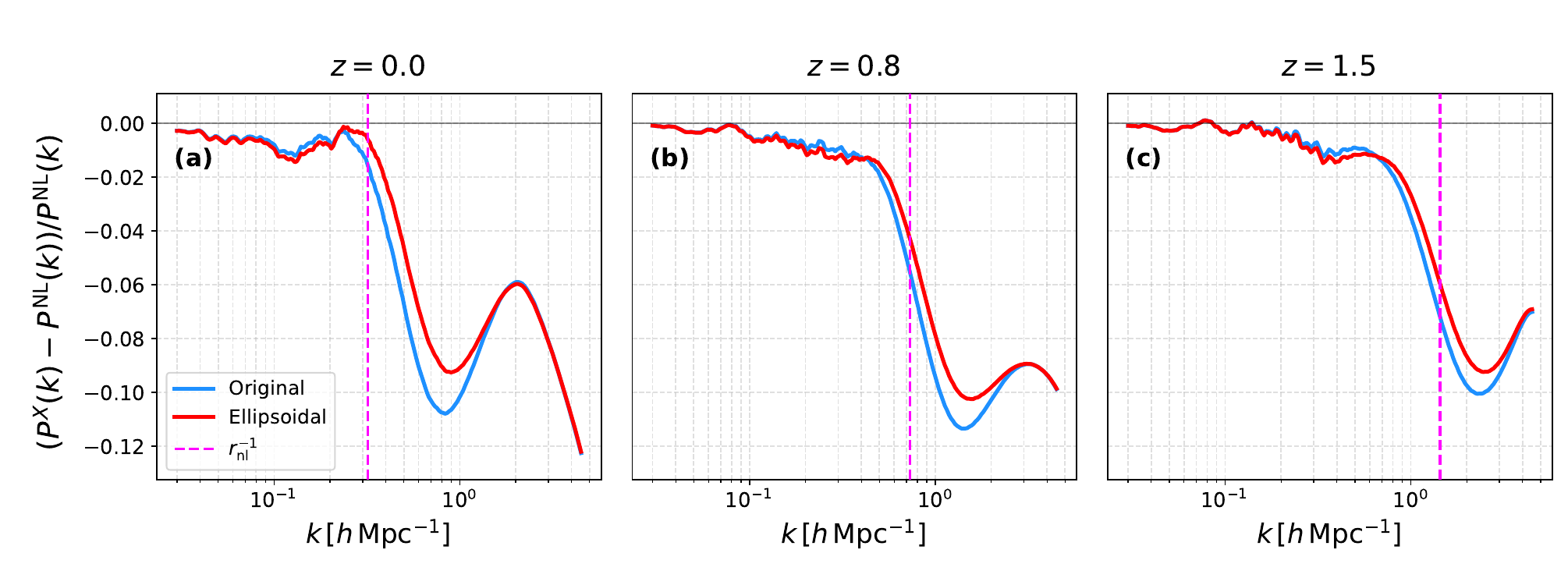}
\caption{
Comparison between the original WHM and the ellipsoidal-window WHM in GR. The three panels correspond to $z=0.0$, $0.8$, and $1.5$. The vertical axis shows the residual with respect to the nonlinear reference spectrum, $(P^{X}-P^{\rm NL})/P^{\rm NL}$, where $P^{\rm NL}$ is the GR nonlinear matter power spectrum from the BACCO emulator \citep{Angulo_2021}, and $P^{X}$ denotes either the original WHM prediction or the ellipsoidal-window WHM prediction. The vertical dashed line marks the characteristic nonlinear scale $k_{\rm nl}=r_{\rm nl}^{-1}$ defined by equation~\eqref{eq:rnl_definition}.}
\label{fig:ellipsoidal_window_results}
\end{figure*}

Figure~\ref{fig:ellipsoidal_window_results} shows that replacing the cylindrical sheet and filament windows by ellipsoidal windows leaves the large-scale behaviour essentially unchanged. This agreement follows from the compensated construction of the WHM. All window functions approach unity as $k\rightarrow0$, so the one-object corrections vanish and the large-scale limit is controlled by the unchanged perturbative contribution $P_{\rm PT}$. The geometrical replacement therefore cannot generate an independent large-scale offset.

The effect of the ellipsoidal windows is concentrated mainly in the transition regime around the nonlinear scale. In all three panels, the original WHM develops a negative residual in this region, indicating that the compensated hierarchy provides too little net power across the transition between successive collapse orders. The ellipsoidal-window prediction is systematically higher in this range and partially fills this dip, although it does not remove the discrepancy completely.

This change can be understood from the role of the window functions in the compensated WHM hierarchy. The ellipsoidal construction preserves the mass and effective overdensity assigned to each collapse order, so the increase in power does not arise from adding matter or changing the abundance of structures. Instead, it changes how the same mass is distributed over Fourier modes. For a collapse order $X=s,f$, the geometrical dependence becomes important when the ellipsoidal argument
\begin{equation}
Q_X(k,\mu)=k\sqrt{a_{\perp,X}^2(1-\mu^2)+a_{\parallel,X}^2\mu^2}
\end{equation}
is of order unity. Depending on the orientation, this corresponds approximately to $ka_{\perp,X}\sim1$ or $ka_{\parallel,X}\sim1$. For $Q_X\ll1$, the normalized window is close to unity and is largely insensitive to the detailed geometry. Around $Q_X\sim1$, however, the cylindrical and ellipsoidal windows distribute the same enclosed mass differently among Fourier modes. Moreover, $W_{\rm s}$ enters with opposite signs in the sheet and filament corrections, while $W_{\rm f}$ enters with opposite signs in the filament and halo corrections. Changing the scale dependence of these windows therefore redistributes compensated power between adjacent collapse orders. In the present calculation, the net redistribution supplies additional power in the range where the original WHM produces its transition-region deficit.

The residual minimum does not coincide exactly with $k_{\rm nl}$. The scale $k_{\rm nl}$ is only a characteristic marker, whereas the WHM integrates over a broad distribution of Lagrangian radii. Even at fixed radius, the orientation-averaged ellipsoidal window changes over a finite wavenumber interval set approximately by $ka_{\perp,X}\sim1$ and $ka_{\parallel,X}\sim1$, rather than at a single value of $k$. The superposition of different mass scales and axis lengths therefore broadens the transition beyond $k_{\rm nl}$.

As the redshift increases, the nonlinear scale moves toward larger wavenumbers. At fixed Lagrangian radius, the linear variance is smaller at earlier times whereas the GR collapse threshold varies only weakly, so the condition $\sigma(r_{\rm nl},z)=\delta_{\rm c}(z)$ is reached at a smaller $r_{\rm nl}$. The transition dip and the effect of the ellipsoidal windows consequently shift toward higher $k$, as seen from $z=0$ to $z=1.5$. The persistence of the improvement over these three redshifts indicates that it is associated with the geometrical structure of the WHM hierarchy rather than with an accidental correction at a single epoch.

At smaller scales, the two curves again become qualitatively similar. This is also expected, since the spherical NFW halo window and the concentration--mass relation are unchanged, while the relative importance of the sheet and filament stages decreases as the halo-order contribution becomes dominant. The ellipsoidal replacement therefore modifies mainly the transition between collapse orders without changing the asymptotic large-scale construction or the adopted small-scale halo structure.

The ellipsoidal windows thus provide a modest but systematic reduction of the transition-region deficit. The remaining negative residual shows that the geometrical replacement alone is not sufficient to reproduce the nonlinear reference spectrum over the full range shown. Nevertheless, the result demonstrates that the finite geometry assigned to sheets and filaments affects the transfer of power between successive collapse stages. We therefore adopt the ellipsoidal windows as the GR baseline for the modified-gravity calculations below.

\subsection{Matter power spectrum boost in \texorpdfstring{$f(R)$}{f(R)} gravity}
\label{sec:pk_boost}

Having constructed the environment-dependent $f(R)$ WHM, we now examine its prediction for the nonlinear matter power spectrum. Rather than comparing the absolute power spectrum, we focus on the boost relative to GR. This ratio isolates the modified-gravity response and reduces the sensitivity of the comparison to inaccuracies shared by the GR and $f(R)$ WHM predictions. It is also the quantity provided directly by the e-MANTIS emulator.

The WHM boost is defined in equation~\eqref{eq:WHM_boost}. We denote the corresponding e-MANTIS prediction by $B_{\rm emu}(k,z)$. Unless otherwise stated, we consider the Hu--Sawicki models with $|f_{R0}|=10^{-5}$ and $10^{-6}$ at $z=0$, $0.8$, and $1.5$. As described in Section~\ref{sec:fR_WHM}, the perturbative contribution $P_{\rm PT}$ and the variance $\sigma(R,z)$ are retained at their GR values. The WHM boost shown below is therefore generated by the modified collapse quantities $\delta_{\rm c}^{f(R)}$ and $\Delta_{\rm v}^{f(R)}$ and their propagation through the compensated sheet, filament, and halo corrections.

\begin{figure*}
\centering
\includegraphics[width=\textwidth]{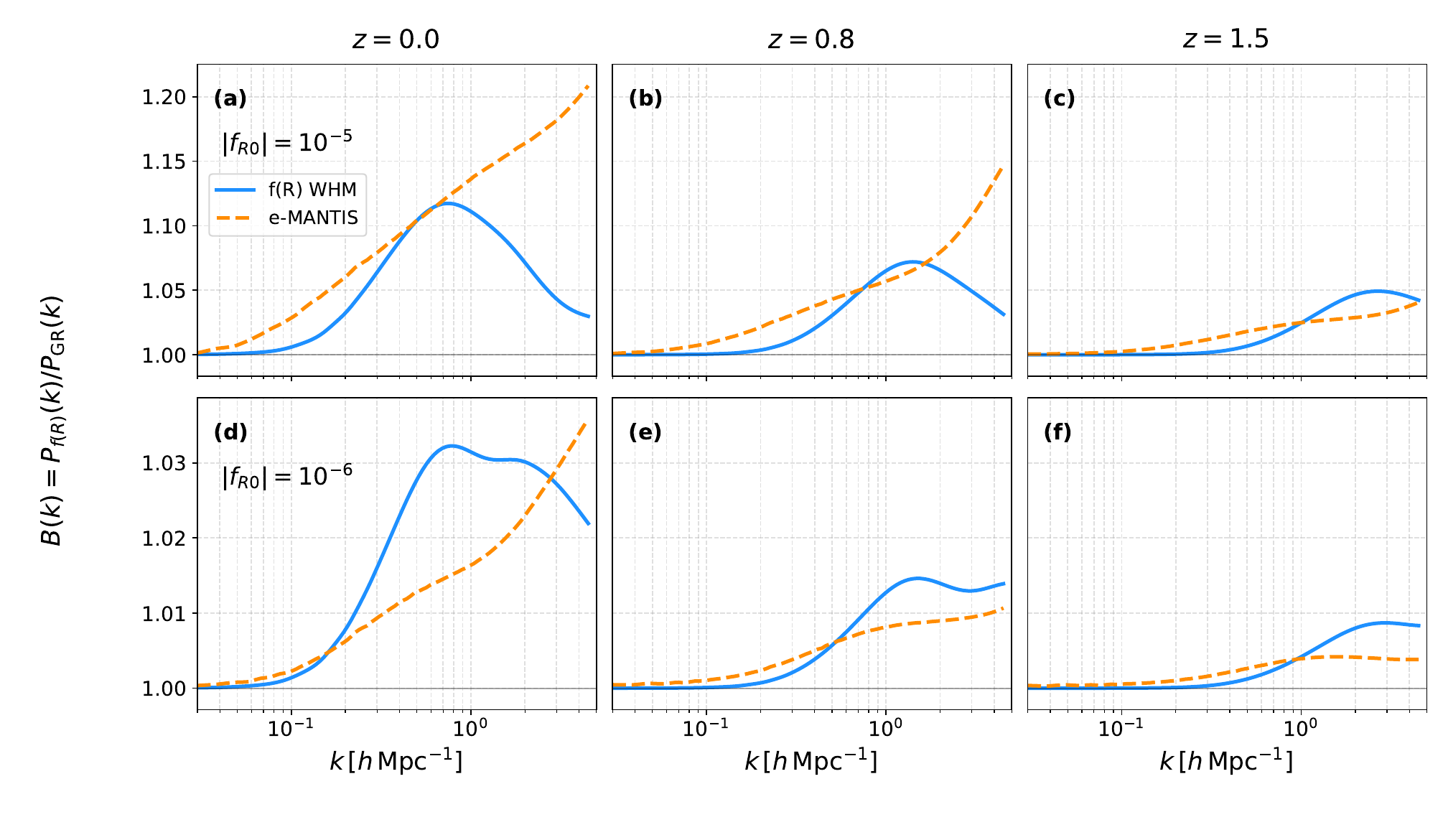}
\caption{
Matter power spectrum boost predicted by the modified WHM and compared with the e-MANTIS emulator. The three columns correspond to $z=0.0$, $0.8$, and $1.5$, while the upper and lower rows show $|f_{R0}|=10^{-5}$ and $10^{-6}$, respectively. The solid blue curves denote the modified WHM prediction, and the dashed orange curves show the e-MANTIS result. Different vertical ranges are used for the two rows in order to display clearly the weaker modified-gravity signal for $|f_{R0}|=10^{-6}$, while the vertical range is kept identical within each row.
}
\label{fig:boost_comparison}
\end{figure*}

Figure~\ref{fig:boost_comparison} shows the boost predicted by the modified WHM and its comparison with e-MANTIS. On large scales, the WHM boost approaches unity for both field strengths and redshifts considered. This behaviour follows directly from our choice of keeping the perturbative contribution $P_{\rm PT}$ fixed to its GR prediction: as the compensated one-object corrections vanish toward small $k$, the numerator and denominator reduce to the same perturbative spectrum. The large-scale return to unity should therefore be interpreted as a boundary of the present collapse-based extension, rather than as a general statement that the full linear response of $f(R)$ gravity is identical to GR.

The modified-gravity signal develops mainly in the quasi-nonlinear and nonlinear regimes, where the sheet, filament, and halo corrections become important. Its amplitude is larger for $|f_{R0}|=10^{-5}$ than for $10^{-6}$. At fixed mass, redshift, and environment, the thin-shell estimator increases with $|f_{R0}|$. A larger field amplitude therefore increases $\Delta R/R_{\rm TH}$ and drives the force enhancement toward its unscreened limit $F=1/3$. Reducing $|f_{R0}|$ instead makes the thin shell narrower, strengthens chameleon screening, and suppresses the power-spectrum response.

The redshift evolution reflects two competing effects in chameleon screening. We denote the background scalar-field value by $\bar f_R(a)\equiv f_R[\bar R(a)]$. For the $n=1$ Hu--Sawicki model, its amplitude scales approximately as $|\bar f_R|\propto\bar R^{-2}$. Toward lower redshift, the decreasing background curvature therefore increases $|\bar f_R|$, which tends to increase the thin-shell factor and make the fifth force less strongly screened. At the same time, nonlinear structures become more developed and their gravitational potentials deepen, which favours self-screening. The power-spectrum boost reflects the net result of these competing trends. For the models and scales considered here, the net redshift evolution follows the background-field trend, with a larger modified-gravity response at lower redshift.

For $|f_{R0}|=10^{-5}$, the modified WHM develops a substantial nonlinear enhancement at all three redshifts. At $z=0$, the WHM and e-MANTIS boosts initially increase in a similar manner and reach comparable amplitudes on intermediate scales, but their small-scale behaviours differ: the WHM response reaches a maximum and then decreases, whereas the emulator boost continues to grow toward larger $k$. A similar, although weaker, turnover appears at $z=0.8$. At $z=1.5$, the overall enhancement is smaller and the WHM remains considerably closer to the emulator over the displayed range, although it still exhibits a somewhat different scale dependence. For $|f_{R0}|=10^{-6}$, the total response is only at the few-per-cent level. At $z=0$, the modified WHM predicts a stronger intermediate-scale enhancement than e-MANTIS, but falls below the emulator at the largest wavenumbers. At $z=0.8$ and $1.5$, both predictions remain close to unity, with the WHM producing a slightly larger enhancement over much of the nonlinear range. 

The intermediate-scale maximum and the subsequent turnover in the WHM should not be interpreted simply as the fifth force becoming weaker above a particular wavenumber. The power spectrum at a given $k$ receives contributions from a broad range of masses and environments, and the modified collapse quantities affect both the multiplicity weights and the structural window scales. The resulting boost is therefore the net response of the compensated sheet, filament, and halo hierarchy. Section~\ref{sec:boost_decomposition} shows explicitly how the relative contributions of these components generate the peak and its subsequent evolution.

To quantify the agreement with the emulator, we define the relative deviation of the modified WHM from e-MANTIS as
\begin{equation}
\epsilon_{\rm WHM}(k,z)=\frac{B_{\rm WHM}(k,z)-B_{\rm emu}(k,z)}{B_{\rm emu}(k,z)}.
\label{eq:relative_boost_difference}
\end{equation}
For reference, we also introduce the GR baseline $B_{\rm GR}(k,z)\equiv 1$, through
\begin{equation}
\epsilon_{\rm GR}(k,z)=\frac{1-B_{\rm emu}(k,z)}{B_{\rm emu}(k,z)}.
\label{eq:GR_boost_difference}
\end{equation}
The quantity $\epsilon_{\rm GR}$ measures the error that would result from neglecting the modified-gravity enhancement entirely.

\begin{figure*}
\centering
\includegraphics[width=\textwidth]{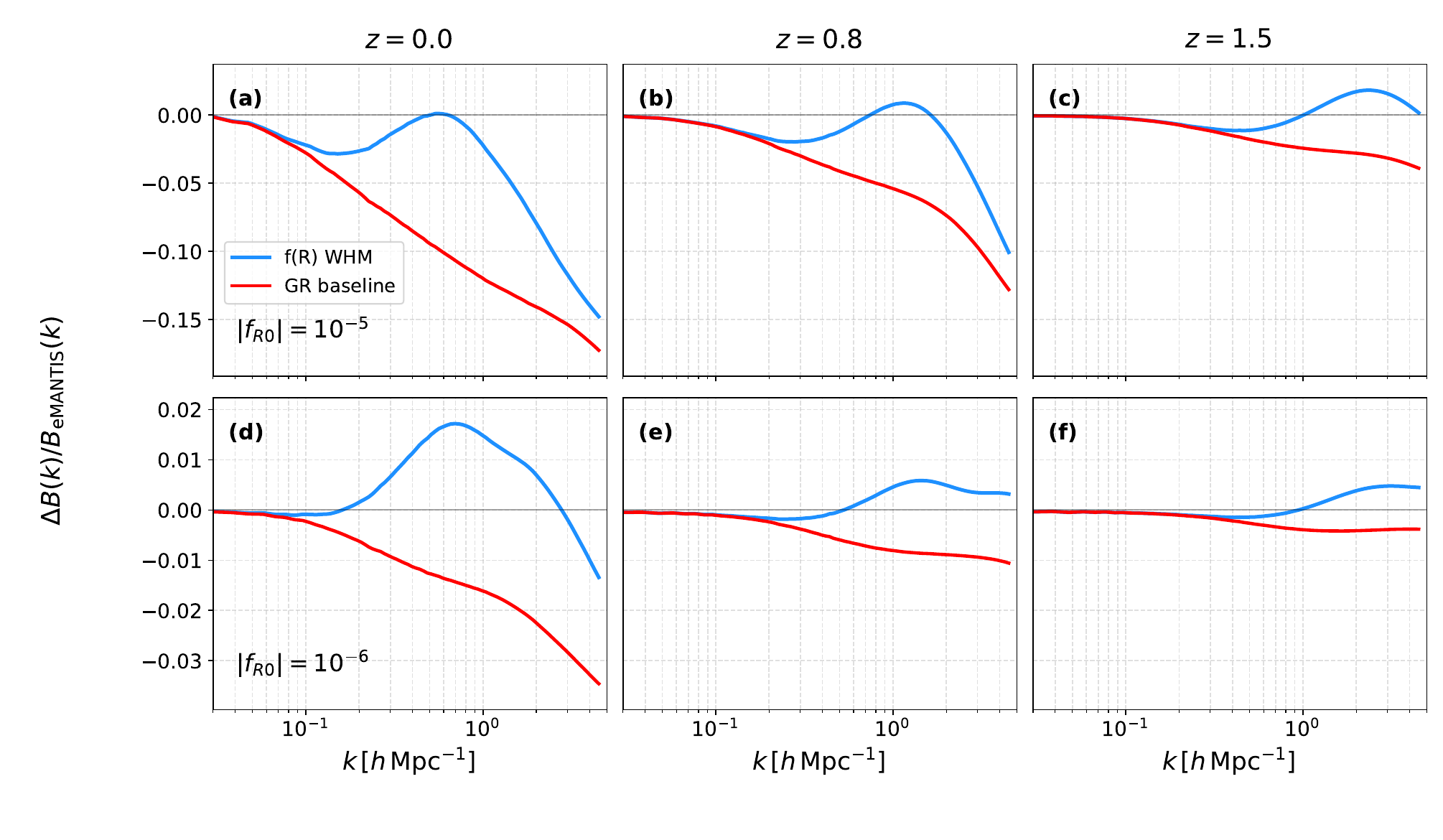}
\caption{
Relative difference with respect to the e-MANTIS boost. The solid blue curves show $\epsilon_{\rm WHM}=(B_{\rm WHM}-B_{\rm emu})/B_{\rm emu}$, while the solid red curves show the GR baseline $\epsilon_{\rm GR}=(1-B_{\rm emu})/B_{\rm emu}$. The three columns correspond to $z=0.0$, $0.8$, and $1.5$, and the upper and lower rows correspond to $|f_{R0}|=10^{-5}$ and $10^{-6}$, respectively. Different vertical ranges are used for the two rows, while the range is kept identical within each row.
}
\label{fig:relative_boost_difference}
\end{figure*}

Figure~\ref{fig:relative_boost_difference} shows that the modified WHM generally provides a closer approximation to the e-MANTIS result than the GR baseline. The improvement is particularly clear on intermediate scales, where the WHM generates a nonzero enhancement with approximately the correct order of magnitude. For $|f_{R0}|=10^{-6}$, the residual remains within approximately two per cent over the full displayed range and is smaller still at $z=0.8$ and $1.5$. A similarly small residual is obtained for $|f_{R0}|=10^{-5}$ at $z=1.5$. The largest discrepancies occur for $|f_{R0}|=10^{-5}$ at $z=0$ and $0.8$. Once the WHM boost passes its intermediate-scale maximum, its residual becomes increasingly negative, reaching its largest magnitude at the highest wavenumbers shown. The modified collapse calculation therefore captures a significant fraction of the $f(R)$ signal in the transition regime but does not reproduce the continued small-scale growth present in the emulator.

This scale dependence is physically informative. The high-$k$ power spectrum is increasingly sensitive to halo abundances and to the internal distribution of matter within haloes. In the present model, modified gravity changes the collapse threshold and virial overdensity, but the functional forms of the multiplicity functions, the NFW profile, and the concentration--mass relation remain inherited from the GR WHM. Full $f(R)$ simulations naturally allow these nonlinear halo properties to respond to modified gravity. The absence of an explicit modified-gravity response in these ingredients is therefore a plausible source of the high-$k$ discrepancy, particularly because halo concentrations and internal profiles directly affect the small-scale power. The approximations involved in the real-space screening, collapse, and virialization prescriptions may also contribute. 

However, the high-$k$ discrepancy should not be attributed solely to the GR ingredients retained in the present $f(R)$ extension. The original WHM already shows a deficit of absolute power on small scales \citep{Brieden_2026}, indicating that its description of the halo-dominated regime is itself incomplete. Taking the ratio to the GR WHM prediction suppresses errors that are common to the numerator and denominator, but does not guarantee their complete cancellation once modified gravity changes the collapse weights and structural scales.

The comparison shows that the collapse-based extension recovers much of the intermediate-scale $f(R)$ response and its ordering with field strength and redshift, while the low-redshift high-$k$ turnover remains the principal discrepancy. To identify the origin of this behaviour within the WHM hierarchy, we next decompose the response into its sheet, filament, and halo contributions.

\subsection{Decomposition of the modified-gravity boost}
\label{sec:boost_decomposition}

The total boost alone does not reveal which part of the modified WHM is responsible for the intermediate-scale enhancement and its subsequent small-scale evolution. We therefore decompose the modified-gravity response into the compensated sheet, filament, and halo corrections. Since the perturbative contribution is kept identical in the $f(R)$ and GR calculations, the entire difference between the two power spectra arises from changes in these three terms.

For each component, we define
\begin{equation}
\Delta P_X(k,z)=\left\langle\hat{P}_{1X}^{f(R)}(k,z)\right\rangle-\hat{P}_{1X}^{\rm GR}(k,z), \qquad X\in\{\mathrm{s},\mathrm{f},\mathrm{h}\},
\label{eq:component_power_difference}
\end{equation}
where the angle brackets denote the environmental average. We then normalize each component difference by the total GR WHM power spectrum,
\begin{equation}
\mathcal{C}_X(k,z)=\frac{\Delta P_X(k,z)}{P_{\rm WHM}^{\rm GR}(k,z)}.
\label{eq:normalized_component_contribution}
\end{equation}
Because $P_{\rm PT}$ is unchanged, these quantities satisfy
\begin{equation}
B_{\rm WHM}(k,z)-1=\mathcal{C}_{\rm s}(k,z)+\mathcal{C}_{\rm f}(k,z)+\mathcal{C}_{\rm h}(k,z).
\label{eq:boost_component_sum}
\end{equation}
We have verified numerically that equation~\eqref{eq:boost_component_sum} is satisfied over the full range of scales considered.

\begin{figure*}
\centering
\includegraphics[width=\textwidth]{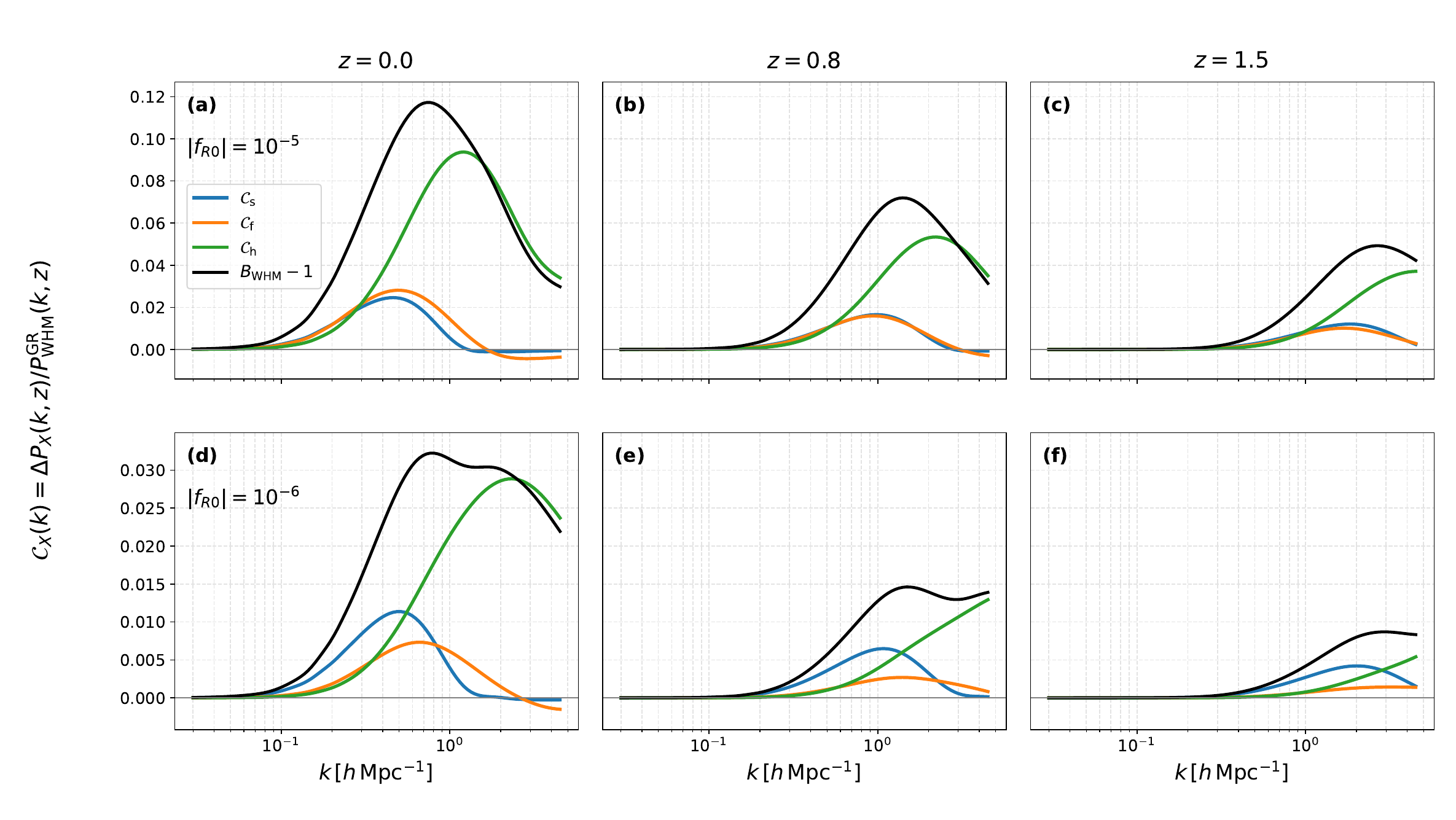}
\caption{
Decomposition of the modified-gravity boost into sheet, filament, and halo responses. The three columns correspond to $z=0.0$, $0.8$, and $1.5$, while the upper and lower rows show $|f_{R0}|=10^{-5}$ and $10^{-6}$, respectively. The blue, orange, and green curves denote $\mathcal{C}_{\rm s}$, $\mathcal{C}_{\rm f}$, and $\mathcal{C}_{\rm h}$, while the black curve shows the total response $B_{\rm WHM}-1$. Different vertical ranges are used for the two rows, while the range is kept identical within each row. A negative value of $\mathcal{C}_X$ means that the corresponding compensated WHM correction is smaller in $f(R)$ than in GR; it does not imply a negative physical power spectrum.}
\label{fig:component_contributions}
\end{figure*}

Figure~\ref{fig:component_contributions} shows that all three normalized component responses vanish on large scales. Their sum therefore approaches zero and gives $B_{\rm WHM}\rightarrow1$, consistently with the large-scale behaviour discussed in Section~\ref{sec:pk_boost}. The modified-gravity response becomes appreciable only when the compensated sheet, filament, and halo corrections begin to contribute.

The component responses have two coupled physical origins. First, the modified collapse threshold changes the peak height $\nu=\delta_{\rm c}^{f(R)}/\sigma_{\rm GR}$, and hence changes the multiplicity weighting assigned to the different collapse orders. Second, the modified virial overdensity changes the characteristic structural radii and therefore the scale dependence of the sheet, filament, and halo windows. Each curve in Figure~\ref{fig:component_contributions} consequently combines a change in collapse weighting with a change in the distribution of power over Fourier modes.

A common hierarchy is visible across the six panels. The sheet and filament responses become important first over similar transition scales and jointly shape the onset of the modified-gravity enhancement, while the halo response reaches its largest values at higher wavenumbers. This ordering reflects the characteristic sizes of the three collapse stages. For the same Lagrangian radius, a sheet retains two axes close to the Lagrangian scale and a filament retains one, whereas a halo has collapsed along all three directions and is therefore more compact. The sheet and filament windows consequently depart from their large-scale limits at smaller $k$, while the halo window continues to affect the response toward smaller spatial scales. 

For $|f_{R0}|=10^{-6}$, the sheet response generally reaches a larger maximum at a slightly lower wavenumber than the filament response. For $|f_{R0}|=10^{-5}$, their amplitudes are more comparable, and the filament response can be slightly larger. This shows that reducing $|f_{R0}|$ does not merely rescale every WHM component by the same factor. Because chameleon screening depends nonlinearly on mass and environment, it also changes the relative response of the different collapse orders.

The subsequent decline of the sheet and filament curves is also a consequence of the compensated hierarchy. The window $W_{\rm s}$ enters the sheet and filament corrections with opposite signs, while $W_{\rm f}$ enters the filament and halo corrections with opposite signs. Changes in these windows therefore redistribute the response between adjacent collapse orders. At sufficiently large $k$, $\mathcal{C}_{\rm s}$ and $\mathcal{C}_{\rm f}$ can become negative, meaning that the corresponding compensated correction is smaller than in GR. This sign change should not be interpreted as a negative power carried by physical sheets or filaments.

The intermediate-scale maximum of the total boost is produced by the combined evolution of the three components rather than by the halo term alone. In several panels, $B_{\rm WHM}-1$ reaches its maximum while $\mathcal{C}_{\rm h}$ is still increasing. The total response turns over when the decrease in $\mathcal{C}_{\rm s}+\mathcal{C}_{\rm f}$ becomes larger than the continuing growth of $\mathcal{C}_{\rm h}$. The position and shape of the boost maximum therefore reflect the changing balance among the sheet, filament, and halo responses across the transition regime.

This component interplay is most apparent at low redshift. For both field strengths at $z=0$, the sheet and filament responses peak on intermediate scales and then decline rapidly. The filament response becomes negative at the largest wavenumbers shown, while the sheet response approaches zero or becomes slightly negative. The halo response remains positive and supplies most of the residual small-scale enhancement, but it also eventually turns over. Their combination produces the strong small-scale decline of the total response. For $|f_{R0}|=10^{-6}$, the separation between the lower-$k$ web response and the higher-$k$ halo response broadens and flattens the maximum of the total boost.

At higher redshifts, the component responses are weaker and their characteristic features move toward larger wavenumbers. This shift is consistent with the evolution of the nonlinear scale: at earlier times, nonlinear collapse is associated with smaller length scales and hence larger $k$. For $|f_{R0}|=10^{-5}$ at $z=0.8$, the halo response eventually turns over and the declining web terms generate a milder version of the low-redshift decrease. At $z=1.5$, the sheet and filament responses remain comparatively small and the halo response continues to dominate the high-$k$ signal, so the turnover of the total boost is much weaker. For $|f_{R0}|=10^{-6}$ at $z=0.8$, the halo response continues to increase over most of the displayed range after the sheet and filament terms have passed their maxima. The growing halo contribution then balances the declining web responses, producing the flattening and weak upturn of the total curve. At $z=1.5$, the same hierarchy is present at a smaller amplitude: the sheet and filament terms shape the initial rise, while the halo term sustains the response toward the largest wavenumbers.

The decomposition shows that the intermediate-scale structure of the modified WHM boost is not simply a halo-only effect. The sheet and filament terms determine how the enhancement first emerges across the transition regime, whereas the halo term controls most of the response at the largest wavenumbers. This component-level picture explains the characteristic scale dependence found in Section~\ref{sec:pk_boost}.

\subsection{BAO diagnostics}
\label{sec:bao_diagnostics}

The nonlinear matter power spectrum is the primary output of the WHM, but its implications can also be examined directly in configuration space through the two-point correlation function. Since $\xi(r)$ is obtained from $P(k)$ by a Fourier transform, the BAO oscillations in the power spectrum generate a localized acoustic feature in the correlation function. The BAO peak therefore provides a natural configuration-space test of how the scale-dependent $f(R)$ response predicted by the WHM propagates beyond the power spectrum itself.

This test is complementary to the direct boost comparison. While the modified-gravity enhancement discussed above is most visible on relatively small scales, the BAO feature probes larger and more weakly nonlinear separations. We therefore examine whether the nonlinear modified-WHM corrections produce any appreciable change in the position or shape of the acoustic feature.

The real-space correlation function is computed from the matter power spectrum as
\begin{equation}
\xi(r,z)=\int_0^\infty\frac{k^2\mathrm{d}k}{2\pi^2}P(k,z)j_0(kr),
\label{eq:xi_from_pk}
\end{equation}
where $j_0$ is the zeroth-order spherical Bessel function. To isolate the oscillatory BAO feature from the broadband correlation function, we subtract a smooth no-wiggle component,
\begin{equation}
\xi_{\rm BAO}(r,z)=\xi(r,z)-\xi_{\rm nw}(r,z).
\label{eq:xi_BAO}
\end{equation}
For the GR and $f(R)$ spectra separately, the corresponding no-wiggle spectrum $P_{\rm nw}(k,z)$ is obtained by Gaussian smoothing of $\ln P$ as a function of $\ln k$. The no-wiggle correlation function $\xi_{\rm nw}(r,z)$ is then computed from $P_{\rm nw}(k,z)$ using the same transform as in equation~\eqref{eq:xi_from_pk}. In the following we use
\begin{equation}
Y_{\rm BAO}(r,z)=r^2\xi_{\rm BAO}(r,z)
\label{eq:Y_BAO}
\end{equation}
to display and characterize the localized acoustic feature. The BAO peak position $r_{\rm peak}$ is determined from the maximum of $Y_{\rm BAO}$ over the acoustic-scale range. The peak amplitude is defined as
\begin{equation}
A_{\rm BAO}=Y_{\rm BAO}(r_{\rm peak}),
\label{eq:bao_amplitude}
\end{equation}
and the peak width $w_{\rm BAO}$ is defined as the full width at half maximum of the same feature.

\begin{figure}
\centering
\includegraphics[width=\columnwidth]{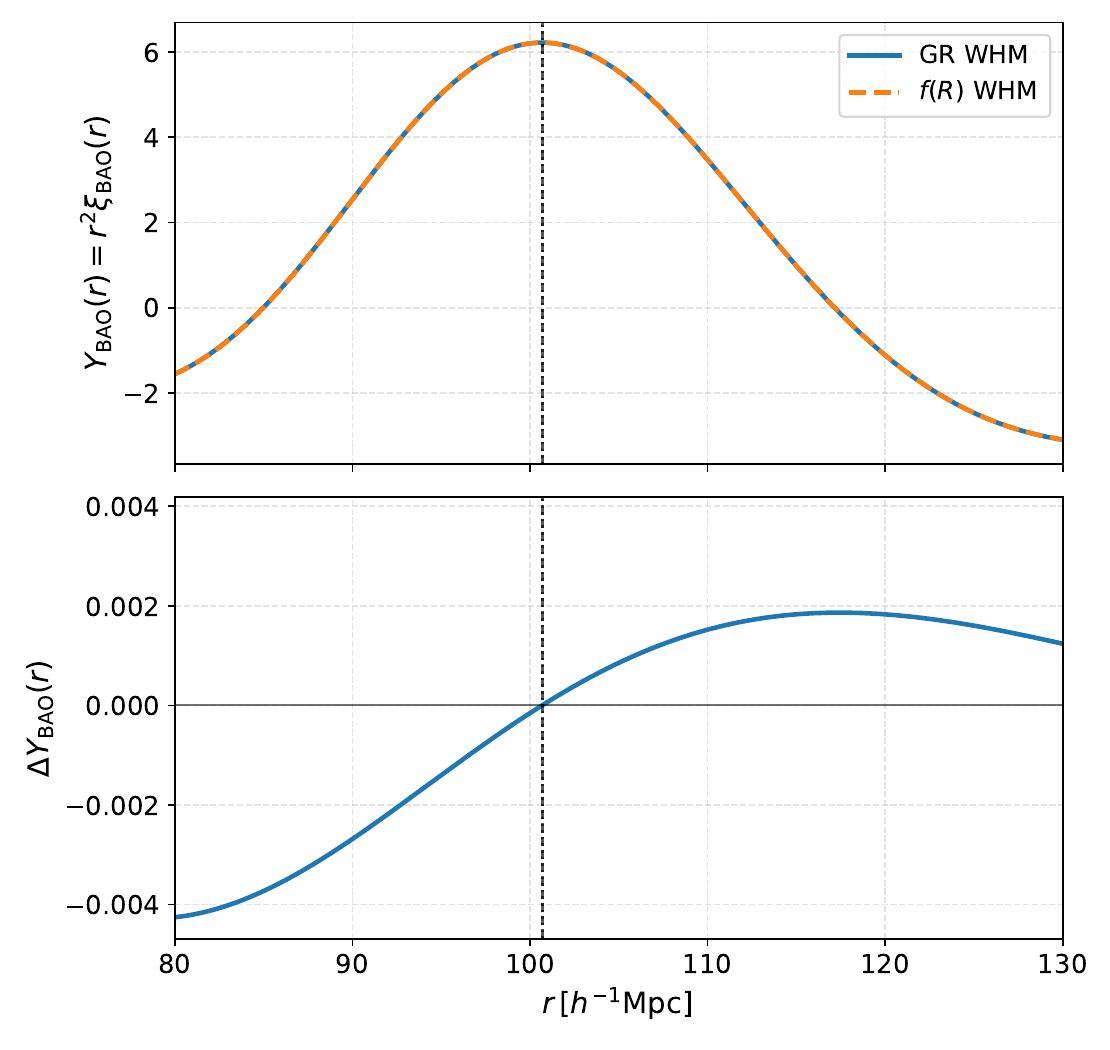}
\caption{
BAO feature for the strongest modified-gravity case considered, $|f_{R0}|=10^{-5}$ at $z=0$. The upper panel shows $Y_{\rm BAO}(r)=r^2\xi_{\rm BAO}(r)$ for GR and the modified WHM prediction, while the lower panel shows $\Delta Y_{\rm BAO}=Y_{\rm BAO}^{f(R)}-Y_{\rm BAO}^{\rm GR}$. The vertical dashed line marks the GR BAO peak position. The two peak profiles almost overlap, and no displacement of the maximum is
resolved on the adopted radial grid.}
\label{fig:bao_peak}
\end{figure}

Figure~\ref{fig:bao_peak} shows the BAO feature for $|f_{R0}|=10^{-5}$ at $z=0$, which gives the largest modified-gravity signal among the cases considered in this work. The GR and $f(R)$ curves are nearly indistinguishable in the upper panel. The lower panel nevertheless reveals a small but nonzero difference around the acoustic feature, showing that the nonlinear corrections slightly reshape the peak profile. The residual is much smaller than the BAO feature itself, and the two maxima occur at the same sampled separation.

A comparison of the peak position alone would not capture this weak change in shape. We characterize the BAO response using the peak position, amplitude, and width for the same field strengths and redshifts considered in the power-spectrum analysis. We define
\begin{equation}
\Delta r_{\rm peak}=r_{\rm peak}^{f(R)}-r_{\rm peak}^{\rm GR},
\label{eq:delta_r_peak}
\end{equation}
and similarly compute the fractional changes in the peak amplitude and width,
\begin{equation}
\frac{\Delta A_{\rm BAO}}{A_{\rm BAO}}=\frac{A_{\rm BAO}^{f(R)}-A_{\rm BAO}^{\rm GR}}{A_{\rm BAO}^{\rm GR}}, \qquad \frac{\Delta w_{\rm BAO}}{w_{\rm BAO}}=\frac{w_{\rm BAO}^{f(R)}-w_{\rm BAO}^{\rm GR}}{w_{\rm BAO}^{\rm GR}
}.
\label{eq:bao_amp_width_changes}
\end{equation}

\begin{table}
\centering
\caption{
BAO peak diagnostics for the modified WHM. The peak position and amplitude are extracted from $Y_{\rm BAO}(r)=r^2\xi_{\rm BAO}(r)$, and the width $w_{\rm BAO}$ is defined as the full width at half maximum of the same peak. Values of $\Delta r_{\rm peak}=0.0$ indicate that the GR and $f(R)$ maxima occur at the same sampled position within the adopted numerical resolution; they should not be interpreted as exact zero shifts.}
\label{tab:bao_diagnostics}
\begin{tabular}{ccccc}
\hline
$|f_{R0}|$ & $z$ &
$\Delta r_{\rm peak}$ [$h^{-1}{\rm Mpc}$] &
$\Delta A_{\rm BAO}/A_{\rm BAO}$ &
$\Delta w_{\rm BAO}/w_{\rm BAO}$ \\
\hline
$10^{-5}$ & $0.0$ & $0.0$ & $1.1\times10^{-6}$ & $-4.5\times10^{-3}$ \\
$10^{-5}$ & $0.8$ & $0.0$ & $-5.7\times10^{-7}$ & $-2.4\times10^{-4}$ \\
$10^{-5}$ & $1.5$ & $0.0$ & $-1.6\times10^{-7}$ & $-2.4\times10^{-5}$ \\
$10^{-6}$ & $0.0$ & $0.0$ & $7.9\times10^{-7}$ & $-1.1\times10^{-3}$ \\
$10^{-6}$ & $0.8$ & $0.0$ & $-1.4\times10^{-7}$ & $-4.9\times10^{-5}$ \\
$10^{-6}$ & $1.5$ & $0.0$ & $-3.5\times10^{-8}$ & $-4.8\times10^{-6}$ \\
\hline
\end{tabular}
\end{table}

The results in Table~\ref{tab:bao_diagnostics} show that no BAO-peak displacement is resolved in any of the cases considered at the adopted numerical resolution. The fractional change in the peak amplitude is at the level of $10^{-6}$ or below. The width is somewhat more sensitive, with the largest change being a narrowing of approximately $0.45$ per cent for $|f_{R0}|=10^{-5}$ at $z=0$; all other width changes are smaller in magnitude. The magnitudes of both fractional changes generally decrease for the weaker field and toward higher redshift, broadly following the reduction of the nonlinear modified-gravity signal found in Section~\ref{sec:pk_boost}.

The weak BAO response can be understood from both the physical origin of the acoustic feature and the structure of the present model. The characteristic acoustic scale is inherited from the early-time matter distribution and the background expansion, which are common to the GR and $f(R)$ calculations considered here. Moreover, the perturbative contribution controlling the large and quasi-linear scales is retained at its GR value. The modified-gravity dependence enters only through the compensated nonlinear sheet, filament, and halo corrections.

These corrections vary relatively smoothly with wavenumber compared with the oscillatory BAO pattern. Their principal effect is therefore a broadband modification of the power spectrum rather than a change in the phase of the acoustic oscillations. The no-wiggle subtraction largely removes this broadband response. In addition, at BAO separations the rapidly oscillating kernel $j_0(kr)$ causes much of the high-$k$ modification to cancel in the Fourier transform. A large nonlinear boost at high $k$ can consequently produce only a very small change in the localized configuration-space BAO feature.

The remaining correction can slightly modify the peak profile, as indicated most clearly by the small change in its width, but it does not produce a resolved shift of the acoustic scale. This explains why the BAO peak position and amplitude are much less sensitive than the nonlinear matter power spectrum itself, while the peak width retains only a small sub-percent response.

This qualitative robustness is consistent with previous modified-gravity studies. In particular, \citet{Lee_2024} found that the BAO linear point, defined as the midpoint between the BAO dip and peak in the correlation function, remains stable in $f(R)$ and DGP models with a $\Lambda$CDM-like background. Although the linear point is not identical to the peak statistic used here, both results indicate that configuration-space acoustic-scale markers are relatively insensitive to late-time modified-gravity evolution. Moreover, compressed BAO analyses have also been found to recover the acoustic scale robustly across broad classes of modified-gravity models \citep{Pan_2024}. 

The small response found here should nevertheless be interpreted within the boundaries of the present WHM construction. In particular, the background expansion and perturbative power spectrum are held fixed between the GR and $f(R)$ calculations. The result therefore shows that the nonlinear collapse corrections included in this work do not produce a resolved BAO-peak shift; it does not establish that every possible linear or quasi-linear $f(R)$ contribution to BAO observables is negligible.

Within this setup, the modified-gravity signal is much more prominent in the nonlinear matter power spectrum than in the BAO peak position. We next turn to weak gravitational lensing, which projects the matter power spectrum over a broad range of nonlinear scales and is therefore expected to retain a more direct response to the modified WHM boost.

\subsection{Weak-lensing angular power spectra}
\label{sec:weak_lensing}

Unlike the BAO peak, which is primarily sensitive to the phase and characteristic scale of the acoustic feature, weak gravitational lensing depends directly on the amplitude of matter clustering projected along the line of sight. The lensing projection receives contributions from a broad range of redshifts and scales, including the nonlinear regime in which the modified WHM boost is largest. It therefore provides a useful projection-level diagnostic of how the predicted $f(R)$ matter-power response propagates into an observable two-point statistic.

We compute tomographic weak-lensing angular power spectra using the Limber approximation \citep{Bartelmann_2001,Kilbinger_2015}, with the standard $\ell+1/2$ mapping \citep{LoVerde_2008},
\begin{equation}
C_\ell^{ij}=\int_0^{z_{\rm lens,max}} \mathrm{d}z\,\frac{\mathrm{d}\chi}{\mathrm{d}z}\frac{W_i(z)W_j(z)}{\chi^2(z)}P\left(k_\ell(z),z\right),
\label{eq:wl_cell}
\end{equation}
where $i$ and $j$ label source-redshift bins and
\begin{equation}
k_\ell(z)=\frac{\ell+1/2}{\chi(z)}
\end{equation}
is the comoving wavenumber selected by the Limber mapping. The integration variable $z$ is the lens redshift, while the source redshift enters through the lensing efficiency kernel. The comoving distance is evaluated in the fiducial flat background,
\begin{equation}
\chi(z)=c\int_0^z\frac{\mathrm{d}z'}{H(z')}.
\end{equation}
For a spatially flat background, the lensing kernel is
\begin{equation}
W_i(z)=\frac{3H_0^2\Omega_m}{2c^2}(1+z)\chi(z)\int_z^{z_{{\rm s},\max}} \mathrm{d}z_s\,n_i(z_s)
\frac{\chi(z_s)-\chi(z)}{\chi(z_s)} .
\label{eq:wl_kernel}
\end{equation}

We adopt an idealized Huterer-type source distribution \citep{Huterer_2002},
\begin{equation}
n(z_s)=\frac{z_s^2}{2z_0^3}\exp\left(-\frac{z_s}{z_0}\right),\qquad z_0=0.5,
\label{eq:wl_source_distribution}
\end{equation}
truncated at $z_{{\rm s},\max}=4.0$ and renormalized over the finite source range. We divide the sources into three tomographic bins with equal source number \citep{Hu_1999}. The bin edges are determined from the cumulative distribution
\begin{equation}
F(z_s)=\frac{\int_0^{z_s}\mathrm{d}z_s'\,n(z_s')}{\int_0^{z_{{\rm s},\max}}\mathrm{d}z_s'\,n(z_s')},
\end{equation}
by requiring $F(z_{s,i})=i/3$ for $i=1,2$. This gives approximately
\begin{equation}
0<z_s<1.01,\qquad 1.01<z_s<1.69,\qquad 1.69<z_s<4.0 .
\end{equation}
Within each bin, the source distribution is normalized according to
\begin{equation}
n_i(z_s)=\frac{n(z_s)\Theta(z_s-z_{i,\min})\Theta(z_{i,\max}-z_s)}{\int_{z_{i,\min}}^{z_{i,\max}}\mathrm{d}z_s'\,n(z_s')}.
\end{equation}
The bin index is ordered by increasing source redshift, so that $C_\ell^{11}$, $C_\ell^{22}$, and $C_\ell^{33}$ denote the auto spectra of the low-, intermediate-, and high-redshift source bins, respectively.

The WHM matter spectra are evaluated at $z=0,0.5,\ldots,3.0$ and interpolated in $(z,\ln k)$. We perform the lensing integral up to $z_{\rm lens,max}=3.0$, corresponding to the highest redshift at which the WHM spectrum is tabulated for this projection. The background distances $\chi(z)$ and geometric lensing kernels $W_i(z)$ are kept fixed to their GR values, consistently with the common fiducial expansion history adopted in the two calculations. We also neglect the direct $f(R)$ correction to the lensing potential, which would otherwise introduce an additional scale- and redshift-dependent factor multiplying $P(k,z)$ in equation~\eqref{eq:wl_cell}. In the regime $|f_R|\ll1$ considered here, the corresponding factor $(1+f_R)^{-1}$ differs negligibly from unity \citep{Lombriser_2013}. Consequently, the geometry and direct lensing response are unchanged between the two calculations, and the difference in $C_\ell^{ij}$ is generated solely by the modified matter power spectrum. 

We do not include shape noise, photometric-redshift uncertainty, intrinsic alignments, or survey window effects \citep{Kilbinger_2015}. Together with the fixed geometry and direct lensing response, these omissions mean that the calculation should be regarded as an idealized matter-power projection rather than a complete modified-gravity weak-lensing forecast.

We characterize the weak-lensing signal through the fractional change
\begin{equation}
R_\ell^{ij}=\frac{C_\ell^{ij,f(R)}}{C_\ell^{ij,\rm GR}}-1 .
\label{eq:wl_ratio}
\end{equation}
Under the approximations adopted here, the GR and $f(R)$ lensing spectra differ only through the matter power spectrum. The fractional response can therefore be written as
\begin{equation}
R_\ell^{ij}=\frac{\displaystyle\int \mathrm{d}z\,\mathcal K_\ell^{ij}(z)P_{\rm WHM}^{\rm GR}\!\left(k_\ell(z),z\right)\left[B_{\rm WHM}\!\left(k_\ell(z),z\right)-1\right]}{\displaystyle\int \mathrm{d}z\,\mathcal K_\ell^{ij}(z)P_{\rm WHM}^{\rm GR}\!\left(k_\ell(z),z\right)},
\label{eq:wl_weighted_boost}
\end{equation}
where
\begin{equation}
\mathcal K_\ell^{ij}(z)=\frac{\mathrm{d}\chi}{\mathrm{d}z}\frac{W_i(z)W_j(z)}{\chi^2(z)}.
\end{equation}
Thus, $R_\ell^{ij}$ is a line-of-sight weighted average of the three-dimensional matter-power response $B_{\rm WHM}(k,z)-1$. A given multipole does not correspond to a single comoving wavenumber or redshift; instead, it combines a range of scales and epochs selected by the lensing kernel.

Since the WHM spectra are tabulated only over a finite $k$-range, we monitor the fraction of the projected signal that samples power outside this range. For each multipole we define
\begin{equation}
f_{\rm clip}^{ij}(\ell)=\frac{\displaystyle\int \mathrm{d}z\, w_\ell^{ij}(z)\Theta\!\left[k_\ell(z)<k_{\min}\ {\rm or}\ k_\ell(z)>k_{\max}\right]}{\displaystyle\int \mathrm{d}z\, w_\ell^{ij}(z)},
\label{eq:wl_clip_fraction}
\end{equation}
where
\begin{equation}
w_\ell^{ij}(z)=\frac{\mathrm{d}\chi}{\mathrm{d}z}\frac{W_i(z)W_j(z)}{\chi^2(z)}P_{\rm WHM}^{\rm GR}\!\left(k_\ell(z),z\right).
\end{equation}
The GR integrand is used as the reference weight so that the reliable multipole range is independent of the modified-gravity amplitude. We regard a multipole as reliable if $f_{\rm clip}^{ij}(\ell)\leq 0.1$. In the following plots, solid segments satisfy this criterion, while dashed segments indicate multipoles for which more than $10\%$ of the projected weight lies outside the tabulated WHM $k$ range.

\begin{figure*}
\centering
\includegraphics[width=\textwidth]{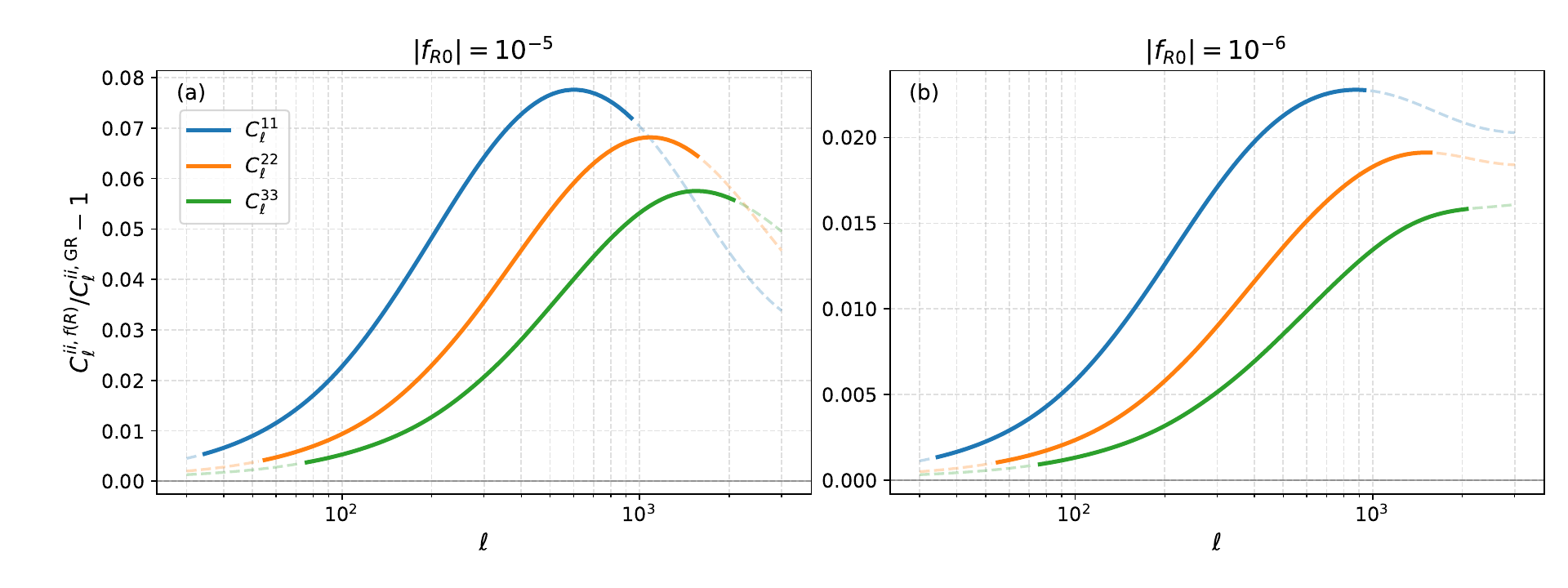}
\caption{
Fractional change of the tomographic weak-lensing auto spectra induced by the modified WHM matter power spectrum. The left and right panels correspond to $|f_{R0}|=10^{-5}$ and $10^{-6}$, respectively. The blue, orange, and green curves show $R_\ell^{ii}\equiv C_\ell^{ii,f(R)}/C_\ell^{ii,\rm GR}-1$ for the low-, intermediate-, and high-redshift source bins. Solid segments indicate multipoles for which no more than $10\%$ of the $P$-weighted Limber integrand lies outside the tabulated WHM $k$ range, while dashed segments mark the corresponding cautionary regions. Different vertical ranges are used in the two panels.}
\label{fig:wl_auto_spectra}
\end{figure*}

Figure~\ref{fig:wl_auto_spectra} shows that the weak-lensing spectra retain a clear percent-level response to the modified WHM matter power spectrum. For $|f_{R0}|=10^{-5}$, the fractional change reaches approximately eight per cent in the lowest source bin and several per cent in the two higher bins. For $|f_{R0}|=10^{-6}$, the response is weaker but still reaches approximately two per cent. This dependence on $|f_{R0}|$ is consistent with the matter power spectrum results: the larger scalar-field amplitude produces a larger nonlinear enhancement, while the weaker model remains closer to the screened GR limit.

The multipole dependence follows from the Limber mapping and from the scale dependence of the underlying boost. At low multipoles, the lensing projection mainly samples relatively large comoving scales, where $B_{\rm WHM}\simeq1$, and the fractional response is consequently small. Increasing $\ell$ moves the weighted projection toward larger wavenumbers, where the nonlinear enhancement becomes appreciable. The response therefore rises as the projection moves into the nonlinear regime and may flatten or turn over once it samples the corresponding feature in the three-dimensional boost.

The line-of-sight integration also smooths this transition, because each multipole combines contributions from a range of redshifts and hence a range of wavenumbers. Features in $B_{\rm WHM}(k,z)$ therefore appear as broader maxima or flattening in $R_\ell^{ij}$. Within the reliable multipole ranges, the later flattening or decline of the lensing response is consistent with the turnover of the underlying modified WHM boost found at low redshift. The dashed portions of the curves should not be interpreted quantitatively because they receive a substantial contribution from outside the tabulated WHM wavenumber range.

The tomographic ordering has a direct geometrical and dynamical interpretation. The lowest source-redshift bin places more weight on late-time lenses, where the $f(R)$ matter-power enhancement is stronger. Moreover, the lower lens redshifts that contribute most strongly to this bin correspond to smaller comoving distances and hence to larger comoving wavenumbers at fixed multipole, $k_\ell=(\ell+1/2)/\chi$. The projection therefore samples both later epochs and more nonlinear scales, increasing the fractional response $R_\ell^{11}$.

The higher source-redshift bins project over a broader range of more distant lenses, including earlier epochs at which the background scalar-field amplitude and the nonlinear matter-power response are smaller. Their fractional changes are therefore reduced. At the same time, a feature at a fixed comoving wavenumber is mapped to a larger angular multipole at larger comoving distance, since $\ell\simeq k\chi$. This explains why the maxima of $R_\ell^{22}$ and $R_\ell^{33}$ occur at progressively larger multipoles than that of $R_\ell^{11}$.

We have also checked the full set of tomographic auto- and cross-spectra and find the same qualitative hierarchy. Only the auto spectra are displayed in Figure~\ref{fig:wl_auto_spectra} for clarity.

The line-of-sight projection smooths, but does not erase, the nonlinear modified-gravity response. Unlike the BAO peak, for which no displacement relative to GR is resolved in the present calculation, the tomographic lensing spectra retain a percent-level imprint of the modified WHM boost. Thus, when the background geometry and direct lensing response are held fixed, the weak-lensing angular spectra are substantially more sensitive to the nonlinear $f(R)$ corrections included in the model.

\section{Discussion and conclusions}
\label{sec:discussion_conclusions}

In this work, we have constructed a semi-analytic extension of the Web--Halo Model to Hu--Sawicki $f(R)$ gravity. The purpose was not to develop a fully calibrated emulator-level prediction, but to retain a direct connection between the nonlinear matter-power response, modified collapse dynamics, and the successive stages of anisotropic cosmic-web formation. The construction combines a geometrical refinement of the intermediate web windows with a dynamical extension of the collapse inputs. The original cylindrical sheet and filament windows were replaced by axisymmetric ellipsoidal windows, while the collapse quantities entering the WHM were generalized to the scale-, redshift-, and environment-dependent functions $\delta_{\rm c}^{f(R)}$ and $\Delta_{\rm v}^{f(R)}$ obtained from a chameleon-screened spherical-collapse calculation.

The ellipsoidal windows produce a modest improvement in the GR matter power spectrum around the transition between perturbative and halo-dominated clustering. The replacement preserves the original WHM mass and overdensity assignments and introduces no additional fitting parameters. Since neither the perturbative contribution nor the NFW halo profile is changed, the large- and small-scale limits remain essentially unchanged. The improvement is concentrated around the nonlinear scale, where the sheet and filament windows control the changing balance between successive collapse orders. This result supports the interpretation that the geometry assigned to intermediate web structures affects how power is redistributed across the transition regime.

The $f(R)$ modification was introduced through the nonlinear collapse equation. The chameleon-screened force changes the collapse trajectory and hence the collapse threshold and virial overdensity. These quantities then modify both the peak heights entering the multiplicity weights and the characteristic radii entering the sheet, filament, and halo windows. The resulting environment-dependent power spectra were averaged over the Eulerian environment distribution. In the present implementation, the variance $\sigma(R,z)$ and the perturbative contribution $P_{\rm PT}$ are retained at their GR values. Accordingly, $B_{\rm WHM}\rightarrow1$ on large scales by construction because the two calculations share the same perturbative sector. This imposed agreement does not imply equality of the full linear $f(R)$ and GR growth responses.

The modified WHM generates an enhancement of the nonlinear matter power spectrum with the expected dependence on field strength and redshift. The response is stronger for $|f_{R0}|=10^{-5}$ than for $10^{-6}$ because the larger scalar-field amplitude produces a larger thin-shell factor and allows the force enhancement to approach its unscreened limit over a wider range of structures. The boost also increases toward lower redshift. This trend is the net result of the evolving background field, object potentials, and environments. Although the growth of nonlinear potential wells favours self-screening at late times, the calculated redshift evolution follows the increasing background-field amplitude toward lower redshift. Compared with the e-MANTIS emulator, the modified WHM generally improves substantially over the GR baseline and captures the onset and approximate magnitude of the nonlinear $f(R)$ enhancement. The agreement is particularly good for the weaker field and at higher redshift. The mismatch is most pronounced for $|f_{R0}|=10^{-5}$ at low redshift and high wavenumber. In these cases, the WHM boost reaches an intermediate-scale maximum and subsequently declines, whereas the emulator exhibits continued growth toward smaller scales.

This high-$k$ discrepancy cannot be assigned to a single modelling ingredient. The present extension modifies $\delta_{\rm c}$ and $\Delta_{\rm v}$ while retaining the GR functional forms of the multiplicity functions, NFW profile, and concentration--mass relation. Modified-gravity changes in these nonlinear halo properties may therefore contribute to the missing small-scale response. At the same time, the original WHM already exhibits a deficit of absolute power in the deeply nonlinear regime. Although the boost ratio suppresses errors common to the GR and modified WHM predictions, changes in the collapse weights and structural scales prevent complete cancellation. Approximations in the real-space screening, collapse, and virialization prescriptions may provide additional contributions, and the present comparison does not allow these effects to be separated uniquely.

The component decomposition attributes the transition-scale structure to the combined web and halo responses. Sheets and filaments govern the initial rise, whereas the halo term becomes increasingly important toward larger $k$; the turnover appears when the declining web contribution overtakes the remaining halo growth. This identifies the boost maximum as a collective feature of the compensated hierarchy rather than a single halo scale.

We propagated the modified WHM spectrum into two configuration- and projection-space diagnostics. No BAO-peak displacement is resolved at the adopted radial
resolution; the amplitude changes remain of order $10^{-6}$, while the width response is sub-percent and reaches approximately $0.45$ per cent in the most affected case. This weak response follows from the common background expansion and perturbative spectrum used in the two calculations. The nonlinear WHM corrections predominantly produce a smooth broadband modification, which is largely removed by the no-wiggle subtraction and strongly suppressed at BAO separations by the oscillatory Fourier kernel. They therefore alter the peak profile only weakly in the present framework. A fully consistent treatment of the linear and quasi-linear $f(R)$ sectors may introduce additional effects.

By contrast, the weak-lensing angular power spectra retain a much clearer imprint of the nonlinear enhancement. The fractional response reaches approximately eight per cent for $|f_{R0}|=10^{-5}$ and approximately two per cent for $10^{-6}$ in the lowest source-redshift bin. The projection averages the three-dimensional response over a broad range of lens redshifts and wavenumbers, smoothing its extrema while preserving a percent-level signal. The fractional response is small at low multipoles, increases as the projection samples increasingly nonlinear wavenumbers, and displays smoother maxima or turnovers than the underlying three-dimensional spectrum. The lowest source-redshift bin has the largest response because it places greater weight on late-time, more nonlinear structures, whereas higher-redshift bins include earlier epochs with a weaker modified-gravity signal and map a fixed comoving wavenumber to larger multipoles. These results describe an idealized projection of the modified matter power spectrum: the background geometry and direct lensing response are fixed, and observational noise and systematic effects are not included.

Several limitations define the scope of the present calculation. First, the linear and quasi-linear sectors are not treated self-consistently in $f(R)$ gravity. Both $P_{\rm PT}$ and $\sigma(R,z)$ are retained at their GR values, so the calculation isolates modified nonlinear collapse rather than the complete scale-dependent growth response. Second, the multiplicity functions, NFW halo profile, and concentration--mass relation are not given explicit modified-gravity dependence, while the original WHM itself remains imperfect in the deeply nonlinear regime. Third, the collapse model uses an approximate real-space thin-shell prescription and a GR $\Lambda$CDM virial-closure relation. The explicit time-dependent fifth-force contribution to the turnaround-to-virialization energy balance is not included. Finally, the BAO and weak-lensing calculations are idealized diagnostics rather than complete observational analyses.

These limitations suggest several extensions for future work. A consistent $f(R)$ perturbative spectrum could replace $P_{\rm PT}$, together with a variance $\sigma(R,z)$ computed from the corresponding modified linear spectrum. This would combine the nonlinear collapse response with modified linear and quasi-linear growth throughout the WHM construction. The multiplicity functions, halo density profile, and concentration--mass relation could be recalibrated using $f(R)$ simulations or supplemented by physically motivated response functions. A more self-consistent virialization treatment would follow the time-dependent chameleon force through the energy balance between turnaround and virialization. Direct comparison of the separate sheet, filament, and halo responses with simulation-based cosmic-web classifications would also provide a more stringent test of the WHM interpretation. The same framework could subsequently be applied to other screened modified-gravity models.

In summary, the WHM can be extended beyond GR while preserving a physically interpretable connection between nonlinear matter clustering and the hierarchy of cosmic-web collapse. The model does not reproduce the emulator response on all nonlinear scales, but it captures the emergence, field-strength dependence, redshift evolution, and approximate amplitude of the $f(R)$ enhancement. The component decomposition further shows how the transition-scale response is distributed among the sheet, filament, and halo collapse orders. Within the present construction, the nonlinear matter power spectrum and tomographic weak-lensing spectra retain the clearest modified-gravity signatures, whereas the BAO peak remains indistinguishable from its GR position at the adopted resolution.

\section*{Acknowledgements}

This work was supported by the National Natural Science Foundation of China under its Key Program (Grant No. 12533002).

During the preparation of this work, the first author used OpenAI's ChatGPT for assistance with code development and debugging, numerical analysis, and language editing. All calculations, interpretations, and manuscript content were reviewed and verified by the authors, who take full responsibility for the work.

%%%%%%%%%%%%%%%%%%%%%%%%%%%%%%%%%%%%%%%%%%%%%%%%%%
\section*{Data Availability}

No new observational data were generated or analysed in this study. The numerical data and analysis scripts underlying the results are available from the corresponding author upon reasonable request.

%%%%%%%%%%%%%%%%%%%% REFERENCES %%%%%%%%%%%%%%%%%%

% The best way to enter references is to use BibTeX:

\bibliographystyle{mnras}
\bibliography{fRWHM} % if your bibtex file is called example.bib

% Alternatively you could enter them by hand, like this:
% This method is tedious and prone to error if you have lots of references
%\begin{thebibliography}{99}
%\bibitem[\protect\citeauthoryear{Author}{2012}]{Author2012}
%Author A.~N., 2013, Journal of Improbable Astronomy, 1, 1
%\bibitem[\protect\citeauthoryear{Others}{2013}]{Others2013}
%Others S., 2012, Journal of Interesting Stuff, 17, 198
%\end{thebibliography}

%%%%%%%%%%%%%%%%%%%%%%%%%%%%%%%%%%%%%%%%%%%%%%%%%%

%%%%%%%%%%%%%%%%% APPENDICES %%%%%%%%%%%%%%%%%%%%%

\appendix

\section{Virialization approximation}
\label{app:virialization}

The computation of the virial overdensity requires a closure relation between the virial and turnaround radii. We define
\begin{equation}
s_{\rm vir}\equiv\frac{r_{\rm vir}}{r_{\rm ta}} .
\end{equation}
In the Einstein--de Sitter limit, the standard spherical-collapse result is $s_{\rm vir}=1/2$. This fixed value is also used in the GR spherical-collapse fitting prescription adopted by \citet{Mead_2016}, and hence in the original WHM implementation. In a $\Lambda$CDM background, however, the cosmological constant contributes to the virialization condition. We therefore determine $s_{\rm vir}$ from the $\Lambda$CDM virial-closure equation rather than fixing it to $1/2$.

We first recall the GR derivation. For a potential-energy contribution with radial dependence
\begin{equation}
U_i \propto r^{n_i},
\end{equation}
the scalar virial theorem gives
\begin{equation}
2T_{\rm vir}=\sum_i n_i U_{i,\rm vir}.
\label{eq:general_virial_theorem}
\end{equation}
For a homogeneous spherical perturbation, the Newtonian gravitational potential energy and the contribution associated with the cosmological constant are
\begin{equation}
U_G=-\frac{3}{5}\frac{GM^2}{r},\qquad U_\Lambda=-\frac{1}{10}\Lambda M r^2 .
\end{equation}
Since $U_G\propto r^{-1}$ and $U_\Lambda\propto r^2$, equation~\eqref{eq:general_virial_theorem} gives
\begin{equation}
T_{\rm vir}=-\frac{1}{2}U_{G,\rm vir}+U_{\Lambda,\rm vir}.
\end{equation}
The total energy at virialization is therefore
\begin{equation}
E_{\rm vir}=T_{\rm vir}+U_{G,\rm vir}+U_{\Lambda,\rm vir}=\frac{1}{2}U_{G,\rm vir}+2U_{\Lambda,\rm vir}.
\end{equation}

At turnaround, the kinetic energy vanishes, so that
\begin{equation}
E_{\rm ta}=U_{G,\rm ta}+U_{\Lambda,\rm ta}.
\end{equation}
Assuming energy conservation between turnaround and virialization, $E_{\rm ta}=E_{\rm vir}$, and writing
\begin{equation}
r_{\rm vir}=s_{\rm vir}r_{\rm ta},
\end{equation}
one obtains \citep{Lahav:1991wc}
\begin{equation}
2\eta_\Lambda s_{\rm vir}^3-(2+\eta_\Lambda)s_{\rm vir}+1=0 ,
\label{eq:virial_s_gr}
\end{equation}
where
\begin{equation}
\eta_\Lambda=\frac{\Lambda}{4\pi G\rho_m(a_{\rm ta})}=2\frac{\Omega_\Lambda}{\Omega_m}
\frac{a_{\rm ta}^3}{1+\delta(a_{\rm ta})}.
\label{eq:eta_lambda}
\end{equation}
For positive $\Lambda$, we choose the physical root that is continuously connected to the Einstein--de Sitter solution, $s_{\rm vir}=1/2$, as $\eta_\Lambda\rightarrow0$.

Equation~\eqref{eq:virial_s_gr} shows that $s_{\rm vir}$ is not a universal constant in a $\Lambda$CDM background. It depends on the turnaround epoch and the density contrast at turnaround through $\eta_\Lambda$. 

In $f(R)$ gravity, a fully self-consistent virialization prescription is more subtle. The fifth-force enhancement is screened by the chameleon mechanism and evolves along the collapse trajectory,
\begin{equation}
G_{\rm eff}=G(1+F),\qquad F=F(R,a,\delta,\delta_{\rm env}^{\rm NL}) .
\end{equation}
The effective gravitational interaction is therefore explicitly time- and environment-dependent. The potential energy of the collapsing matter cannot be represented by a time-independent function of radius alone, and the standard turnaround-to-virialization energy closure cannot be applied without an additional approximation.

To assess the sensitivity of $s_{\rm vir}$ to an explicit fifth-force contribution, we replace the evolving enhancement by a constant effective value $\bar F$ between turnaround and virialization. In this simplified treatment, the Newtonian potential term is rescaled by $1+\bar F$, leading to
\begin{equation}
2\eta_\Lambda s_{\rm vir}^3-\left[2(1+\bar F)+\eta_\Lambda\right]s_{\rm vir}+1+\bar F=0 .
\label{eq:virial_s_fR_effective}
\end{equation}
This equation reduces to equation~\eqref{eq:virial_s_gr} when $\bar F=0$. We emphasize that equation~\eqref{eq:virial_s_fR_effective} is used only as a diagnostic of the sensitivity to an effective fifth-force contribution; it is not adopted as the fiducial virialization model.

In the numerical collapse calculation, different values of $R$, $\delta_{\rm env}$, $z$, and $|f_{R0}|$ produce different turnaround times and overdensities, and therefore different values of $\eta_\Lambda$. Once $\eta_\Lambda$ is fixed, equation~\eqref{eq:virial_s_fR_effective} depends only on $\eta_\Lambda$ and $\bar F$. To isolate the sensitivity to $\bar F$, Table~\ref{tab:svir_sensitivity} uses representative fixed values of $\eta_\Lambda$ selected from the ranges found in the numerical collapse grid. To bracket the sensitivity, we consider $\bar F=1/6$ as an intermediate effective enhancement and $\bar F=1/3$ as the fully unscreened limit.

\begin{table}
\centering
\caption{
Sensitivity of the virial-to-turnaround radius ratio to the effective fifth-force parameter $\bar F$ at fixed $\eta_\Lambda$. The column $\eta_\Lambda^{\rm grid}$ shows the range found in the numerical collapse grid over $|f_{R0}|=10^{-6}$ and $10^{-5}$, $R=0.1,0.3,1,3,10\,h^{-1}{\rm Mpc}$, and $\delta_{\rm env}=-0.5,0,0.5$. The representative value $\eta_\Lambda^{\rm rep}$ is held fixed when solving equation~\eqref{eq:virial_s_fR_effective}. Here we define $\Delta s(\bar F)\equiv s_{\rm vir}(\bar F)-s_{\rm vir}(0)$.}
\label{tab:svir_sensitivity}
\footnotesize
\setlength{\tabcolsep}{2.5pt}
\begin{tabular}{@{}cccccc@{}}
\hline
$z$ &
$\eta_\Lambda^{\rm grid}$ &
$\eta_\Lambda^{\rm rep}$ &
$s_{\rm vir}(0)$ &
$\Delta s(1/6)$ &
$\Delta s(1/3)$ \\
\hline
$0.0$ &
$0.1025$--$0.1260$ &
$0.12$ &
$0.4846$ &
$2.25\times10^{-3}$ &
$3.93\times10^{-3}$ \\
$0.8$ &
$0.0246$--$0.0304$ &
$0.03$ &
$0.4962$ &
$5.43\times10^{-4}$ &
$9.50\times10^{-4}$ \\
$1.5$ &
$0.00977$--$0.0121$ &
$0.01$ &
$0.4987$ &
$1.79\times10^{-4}$ &
$3.14\times10^{-4}$ \\
\hline
\end{tabular}
\end{table}

Table~\ref{tab:svir_sensitivity} illustrates that the representative values of $\eta_\Lambda$ decrease with increasing redshift, so that the $\bar F=0$ solution approaches the Einstein--de Sitter value $s_{\rm vir}=1/2$. At fixed $\eta_\Lambda$, increasing $\bar F$ from $0$ to $1/3$ changes $s_{\rm vir}$ by less than $4\times10^{-3}$ for the low-redshift representative value, and by less than $10^{-3}$ at the two higher redshifts.

Since these shifts are small and the constant-$\bar F$ prescription is only an effective diagnostic rather than a self-consistent treatment of the time-dependent chameleon force, we adopt equation~\eqref{eq:virial_s_gr}, corresponding to $\bar F=0$, as the fiducial closure in the main calculation. This does not amount to fixing $s_{\rm vir}=1/2$: the value of $s_{\rm vir}$ is determined separately for each collapse trajectory through its turnaround quantities and the corresponding $\eta_\Lambda$.

With this closure, the virial overdensity is given by equation~\eqref{eq:Delta_v_fR}. The resulting $\Delta_{\rm v}^{f(R)}$ still inherits its dependence on scale, redshift, environment, and field strength from the modified collapse trajectory through $a_{\rm ta}$, $\delta(a_{\rm ta})$, and $\eta_\Lambda$. The only additional approximation in this closure is the neglect of the explicit time-dependent fifth-force contribution to the turnaround-to-virialization energy balance.

\section{Collapse diagnostics and limitations of the real-space linearization}
\label{app:collapse_diagnostics}

The modified WHM receives its spherical-collapse inputs through the collapse threshold $\delta_{\rm c}^{f(R)}$ and the virial overdensity $\Delta_{\rm v}^{f(R)}$. Because the chameleon-screened force evolves along the collapse trajectory, both quantities depend on Lagrangian radius, collapse redshift, field strength, and environmental density. This appendix documents representative properties of these collapse outputs and examines a limitation of the real-space prescription used to construct the linearized evolution. We first show their scale, redshift, and field-strength dependence at a fixed underdense environment. We then analyse the non-smooth environmental dependence generated by clipping the linearized thin-shell force to its physical range.

\subsection{Behaviour of the collapse quantities}
\label{app:collapse_quantities}

For each combination of $R$, $z$, and $\delta_{\rm env}$, the nonlinear initial overdensity is adjusted so that collapse occurs at
\begin{equation}
a_{\rm coll}=\frac{1}{1+z}.
\end{equation}
The corresponding linear equation is then evolved from the same initial conditions, and the linear density contrast at $a_{\rm coll}$ defines the collapse threshold,
\begin{equation}
\delta_{\rm c}^{f(R)}=\delta_{\rm lin}(a_{\rm coll}).
\end{equation}
The virial overdensity is obtained from the nonlinear turnaround
quantities,
\begin{equation}
\Delta_{\rm v}^{f(R)}=\frac{1}{s_{\rm vir}^3}\left[1+\delta(a_{\rm ta})\right]\left(\frac{a_{\rm coll}}{a_{\rm ta}}\right)^3,
\end{equation}
where $s_{\rm vir}$ is determined using the virialization closure described in Appendix~\ref{app:virialization}.

\begin{figure*}
\centering
\includegraphics[width=\textwidth]{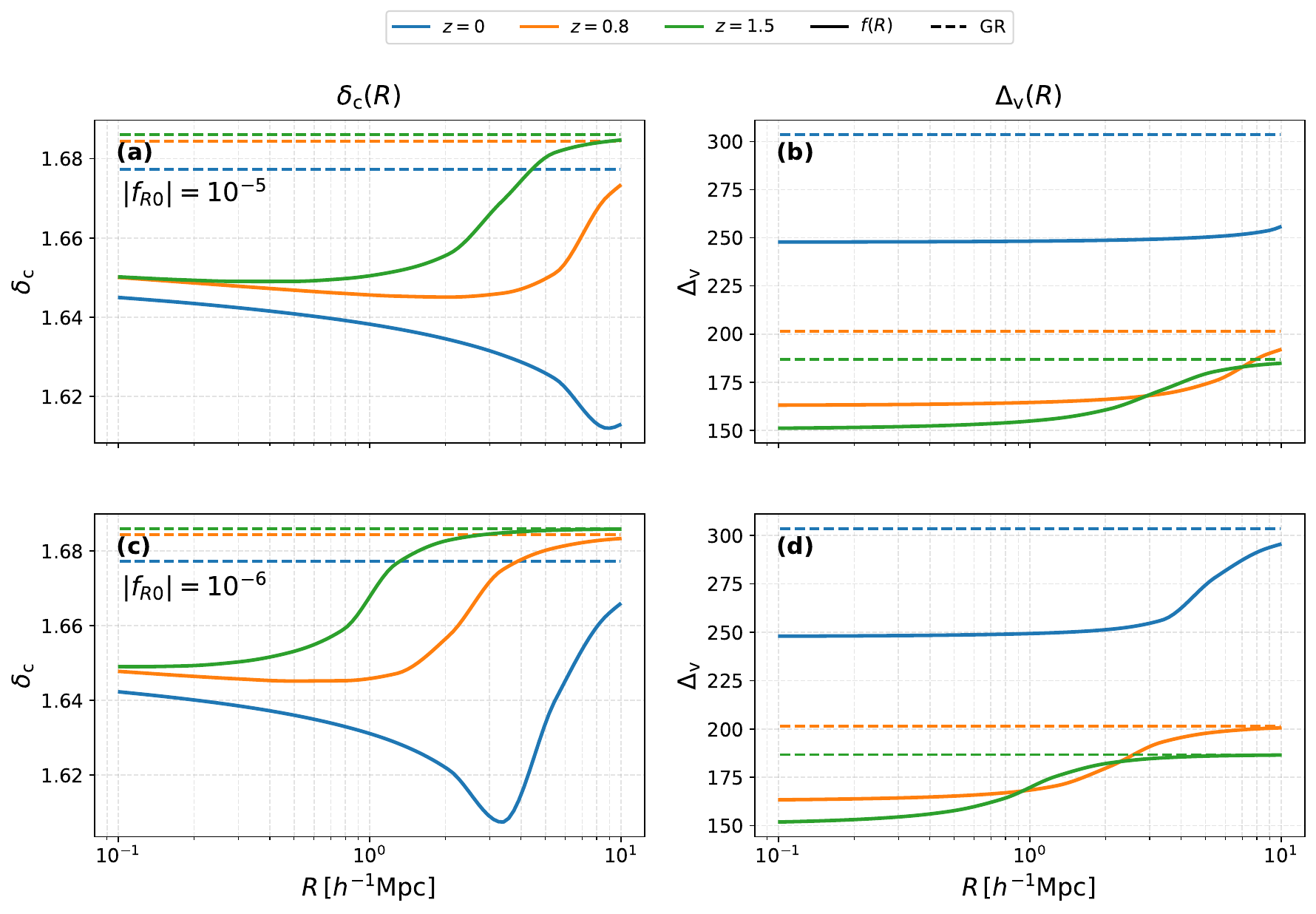}
\caption{
Representative spherical-collapse outputs at the fixed environmental density $\delta_{\rm env}=-0.5$. The left and right columns show the collapse threshold $\delta_{\rm c}^{f(R)}$ and virial overdensity $\Delta_{\rm v}^{f(R)}$, respectively, as functions of the Lagrangian radius. The upper and lower rows correspond to $|f_{R0}|=10^{-5}$ and $10^{-6}$, while the colours denote the collapse redshifts $z=0$, $0.8$, and $1.5$. Solid curves show the $f(R)$ results and dashed curves show the corresponding GR values. The environmental dependence near the clipping boundaries of the linearized force is examined separately in Figure~\ref{fig:linear_force_clipping}.}
\label{fig:collapse_quantities}
\end{figure*}

Figure~\ref{fig:collapse_quantities} shows the scale and redshift dependence of the collapse outputs at the fixed underdense environment $\delta_{\rm env}=-0.5$. This choice avoids the environmental clipping feature examined in Section~\ref{app:linear_force_clipping}, allowing the dependence on $R$, $z$, and $|f_{R0}|$ to be displayed more clearly.

Over the range in which the fifth force affects the collapse trajectory, both $\delta_{\rm c}^{f(R)}$ and $\Delta_{\rm v}^{f(R)}$ depart from their GR values. The collapse threshold is generally reduced because the enhanced gravitational force allows a smaller initial overdensity to reach collapse at the specified redshift. The virial overdensity is also lower than its GR counterpart over most of the modified regime, reflecting the changes in the turnaround epoch, turnaround overdensity, and resulting virial-to-turnaround radius ratio.

The range of affected radii depends strongly on the field amplitude. For $|f_{R0}|=10^{-6}$, the collapse quantities approach their GR values at smaller radii than for $|f_{R0}|=10^{-5}$. The larger field amplitude allows the fifth force to remain effective in structures with deeper gravitational potentials and therefore extends the modified regime to larger Lagrangian radii.

The redshift dependence follows the evolution of the background scalar field and the collapse trajectory. At higher redshift, the background field amplitude is smaller and the deviations from GR are generally confined to smaller radii. The detailed evolution also depends on the object potential and environmental density entering the thin-shell factor, so the curves represent the net response of these effects rather than the variation of any single screening quantity.

The scale dependences of $\delta_{\rm c}^{f(R)}$ and $\Delta_{\rm v}^{f(R)}$ are not identical. The collapse threshold is defined from the linearized evolution associated with the initial condition that produces nonlinear collapse, whereas the virial overdensity is determined directly from the nonlinear turnaround epoch, turnaround overdensity, and virial-to-turnaround radius ratio. Consequently, $\delta_{\rm c}^{f(R)}$ can exhibit a non-monotonic transition with Lagrangian radius, while $\Delta_{\rm v}^{f(R)}$ varies more smoothly and need not recover its GR limit at exactly the same scale.

\subsection{Environmental dependence and force clipping}
\label{app:linear_force_clipping}

The linearized collapse equation is constructed using the same real-space screening prescription as the nonlinear equation, but with the interior density contrast set to zero,
\begin{equation}
F_{\rm lin}=F(R,a,\delta=0,\delta_{\rm env}^{\rm NL}).
\end{equation}
This choice maintains a direct connection between the nonlinear and linearized collapse calculations while keeping the evolution equation linear in $\delta_{\rm lin}$. It is nevertheless a practical real-space prescription rather than a linear response derived directly from the $f(R)$ field equations.

When the interior contrast is set to zero, the normalized densities entering the thin-shell expression become
\begin{equation}
\widetilde{\rho}_{\rm in}=1, \qquad \widetilde{\rho}_{\rm out}=1+\delta_{\rm env}^{\rm NL}.
\end{equation}
For an overdense environment, $\widetilde{\rho}_{\rm out}>\widetilde{\rho}_{\rm in}$, and the formal thin-shell estimate becomes negative. This lies outside the physical thin-shell construction, for which the force enhancement is bounded by $0\leq F_{\rm lin}\leq1/3$, and therefore requires a lower clipping condition.

To display the resulting piecewise structure explicitly, we define the unclipped thin-shell variable
\begin{equation}
\mathcal X_{\rm lin}\equiv 3\alpha \left. \frac{\Delta R}{R_{\rm TH}} \right|_{\delta=0}.
\label{eq:Xlin_definition}
\end{equation}
The force enhancement used in the linearized collapse equation is then
\begin{equation}
F_{\rm lin} =
\begin{cases}
\dfrac{1}{3},&\mathcal X_{\rm lin}\geq1,\\[6pt]
\dfrac{\mathcal X_{\rm lin}}{3},&0<\mathcal X_{\rm lin}<1,\\[6pt]
0,&\mathcal X_{\rm lin}\leq0.
\end{cases}
\label{eq:F_lin_clipping}
\end{equation}
The upper branch corresponds to the saturated, fully unscreened force, while the intermediate branch represents a partially enhanced force. The lower branch enforces the bound $F_{\rm lin}\geq0$ when the formal thin-shell estimate becomes negative. In the present linearization, setting $\delta=0$ fixes $\widetilde{\rho}_{\rm in}=1$. The lower clipping boundary therefore occurs at $\delta_{\rm env}^{\rm NL}=0$, while the zero-force branch applies for $\delta_{\rm env}^{\rm NL}>0$.

\begin{figure}
\centering
\includegraphics[width=\columnwidth]{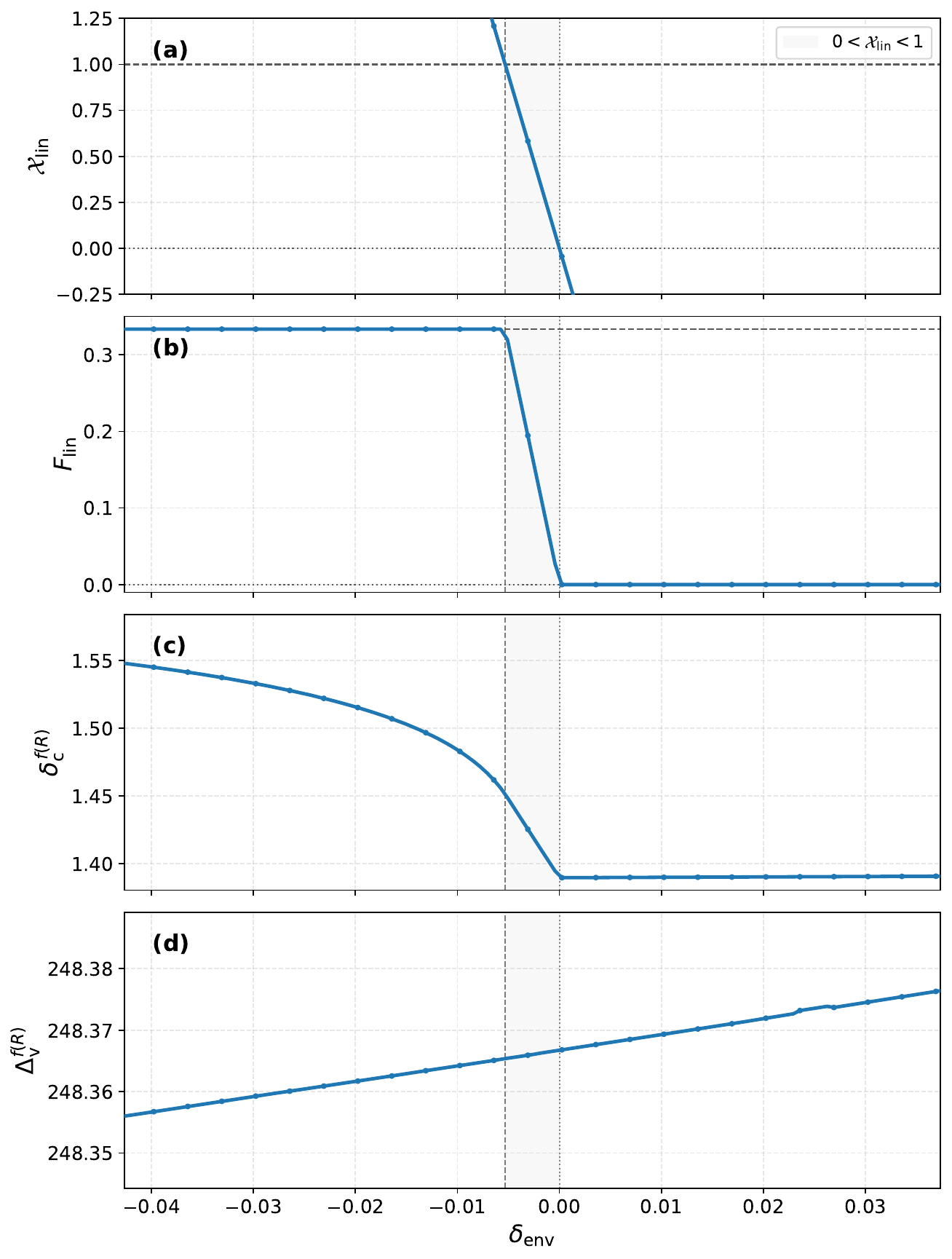}
\caption{
Clipping diagnostic for the real-space linear screening prescription at $|f_{R0}|=10^{-5}$, $z=0$, and $R=1\,h^{-1}{\rm Mpc}$. Panel (a) shows the unclipped thin-shell variable $\mathcal X_{\rm lin}=3\alpha\Delta R/R_{\rm TH}$ evaluated at $a=a_{\rm coll}$, and panel (b) shows the corresponding bounded force enhancement $F_{\rm lin}$ defined by equation~\eqref{eq:F_lin_clipping}. Panels (c) and (d) show the resulting collapse threshold $\delta_{\rm c}^{f(R)}$ and virial overdensity $\Delta_{\rm v}^{f(R)}$, respectively. The vertical dashed and dotted lines mark the environmental values at which $\mathcal X_{\rm lin}=1$ and $\mathcal X_{\rm lin}=0$, and the shaded interval denotes the partially enhanced branch $0<\mathcal X_{\rm lin}<1$. The force quantities are shown at $a=a_{\rm coll}$ to identify the clipping boundaries, whereas $\delta_{\rm c}^{f(R)}$ depends on the full time evolution of $F_{\rm lin}$.
}
\label{fig:linear_force_clipping}
\end{figure}

Figure~\ref{fig:linear_force_clipping} illustrates these branches for $|f_{R0}|=10^{-5}$, $z=0$, and $R=1\,h^{-1}{\rm Mpc}$. The displayed environmental interval is restricted to the neighbourhood of the transition in order to resolve the two clipping boundaries. Panels (a) and (b) show $\mathcal X_{\rm lin}$ and $F_{\rm lin}$ at $a=a_{\rm coll}$. As $\delta_{\rm env}$ increases, $\mathcal X_{\rm lin}$ first crosses the saturation boundary $\mathcal X_{\rm lin}=1$ and then the lower boundary $\mathcal X_{\rm lin}=0$. The force enhancement correspondingly moves from the saturated branch, through the partially enhanced branch, to $F_{\rm lin}=0$.

Panel (c) shows a rapid but continuous transition in $\delta_{\rm c}^{f(R)}$ across the same environmental interval. The threshold decreases as the linearized force moves from the saturated branch to the zero-force branch, after which its environmental dependence becomes nearly flat. The force quantities in panels (a) and (b) are evaluated only at $a=a_{\rm coll}$ in order to identify the clipping boundaries, whereas $\delta_{\rm c}^{f(R)}$ depends on the full time evolution of $F_{\rm lin}$. Their correspondence should therefore not be interpreted as an instantaneous algebraic relation.

Although the clipping interval is narrow, the associated change in $\delta_{\rm c}^{f(R)}$ is appreciable. Its relevance cannot therefore be inferred from the width of the interval alone. The position and width of the transition also depend on $R$, $z$, and $|f_{R0}|$, so the case shown in Figure~\ref{fig:linear_force_clipping} is a representative diagnostic rather than a universal clipping location.

No analogous feature appears in panel (d), where $\Delta_{\rm v}^{f(R)}$ varies smoothly with environmental density. This difference arises because $\Delta_{\rm v}^{f(R)}$ is obtained from the nonlinear collapse trajectory and does not use $F_{\rm lin}$. The rapid transition in $\delta_{\rm c}^{f(R)}$ is therefore specific to the real-space linearization rather than a corresponding irregularity in the nonlinear collapse solution.

The environmental transition in $\delta_{\rm c}^{f(R)}$ should not be interpreted as a physical non-smoothness of linear perturbations in $f(R)$ gravity. It results from applying the nonlinear real-space thin-shell prescription to an interior density fixed at $\delta=0$ while retaining a nonlinear environmental density, followed by clipping the resulting force to its allowed range. The behaviour within the clipping interval should therefore be regarded as a modelling limitation of the linearization adopted in the present semi-analytic framework. A more self-consistent treatment would require deriving the linear response directly from the scalar-field perturbation equations while retaining a suitable connection to the nonlinear screened-collapse calculation.

%%%%%%%%%%%%%%%%%%%%%%%%%%%%%%%%%%%%%%%%%%%%%%%%%%

% Don't change these lines
\bsp	% typesetting comment
\label{lastpage}
\end{document}